\documentclass{article}
\usepackage{PRIMEarxiv}
\usepackage{amsmath,amssymb}
\usepackage[T1]{fontenc}    
\usepackage{hyperref}       
\usepackage{url}            
\usepackage{booktabs}       
\usepackage{amsfonts}       
\usepackage{nicefrac}       
\usepackage{microtype}      
\UseMicrotypeSet[protrusion]{basicmath} 
\usepackage{lipsum}
\usepackage{fancyhdr}       
\usepackage{graphicx}       
\graphicspath{{media/}}     
\usepackage{orcidlink}     
\usepackage{authblk}    
\numberwithin{equation}{section}
\usepackage{setspace}
\title{The Cost of Nonlocality: A Dynamical Performance Equation of Energy–Entanglement–Complexity

}
\author{%
  HongZheng Liu\,\orcidlink{0009-0002-2238-9187}$^{1,*,\dagger}$,
  YiNuo Tian\,\orcidlink{0009-0005-8088-9894}$^{1}$,
  Zhiyue Wu\,\orcidlink{0009-0003-4765-2049}$^{1}$\\
  $^1$Independent, China\\
  $^*$First author\\
  $^\dagger$\textbf{Corresponding author: weiyouyeyu@foxmail.com}%
}
\begin{document}
\maketitle

\begin{abstract}
This work aims to quantify the physical cost of generating non-local entanglement in systems governed by local interactions. 
By unifying the quantum speed limit and Lieb-Robinson bounds, we establish an "energy-entanglement performance equation." 
This framework connects theoretical computational complexity with experimental observables by introducing a measurable proxy for complexity, thereby revealing a performance trade-off among the "energy variance-entanglement product," the strength of local interactions, and dynamical efficiency. 
Our work not only defines a "performance frontier"—constrained by theoretical bounds and amenable to experimental benchmarking—but also provides a novel diagnostic tool for identifying the performance bottlenecks of a process.
\end{abstract}


\section{Introduction}
In the edifice of fundamental physics, profound laws are often expressed as universal inequalities that delineate insurmountable bounds for physical processes. The Quantum Speed Limit (QSL), for instance, establishes a fundamental timescale for quantum dynamics by linking the dynamical resources driving a system's evolution to the maximal rate of that evolution \cite{MandelstamTamm1945, MargolusLevitin1998, DeffnerCampbell2017}. Concurrently, the Lieb-Robinson (LR) bounds strictly constrain the speed of information propagation in systems governed by local interactions, thereby defining an effective spacetime light cone for causal evolution in many-body systems \cite{LiebRobinson1972, hastings2010locality}. These two principles jointly underpin our understanding of quantum dynamics. 
A key question arises when these fundamental principles are considered in the context of the core challenges facing contemporary quantum technology: How is the central resource of quantum computation—non-local quantum entanglement—efficiently generated in a physical system strictly constrained by the principle of locality (as delineated by the LR bounds)? What, precisely, is the quantitative physical cost of establishing such non-local correlations? This question is far from being a matter of purely theoretical speculation. In any real-world quantum computing or information processing task, the dynamical resources available to drive the system's evolution constitute a hard physical constraint. Without a precise resource consumption model that simultaneously accounts for the cost of dynamical resources, locality limitations, and computational complexity, any assessment of a claimed "quantum advantage" derived from entanglement would rest upon an incomplete accounting\cite{BravyiHastingsVerstraete2006,VanAcoleyen2013,Beny2018}. While existing theoretical works, such as those applying QSL or LR bounds, have offered crucial constraints in specific contexts \cite{Eisert2010, BravyiHastingsVerstraete2006}, a theory that unifies the dynamical cost originating from energy uncertainty (QSL) with the information propagation cost stemming from the principle of locality (LR bounds) into a single, self-consistent resource accounting framework—and thereby precisely quantifies the cost of generating non-locality—remains to be developed. This paper aims to systematically fill this theoretical gap. Its core contributions can be summarized in three points:

\begin{enumerate}
    \item Resource Accounting Framework: By introducing a pair of measurable "dynamical efficiency factors," we unify the two fundamental physical bounds into an algebraically exact "energy-entanglement performance equation." This resource identity provides a precise ledger for the "cost of non-locality," quantitatively revealing the profound connection among energy variance, entanglement, complexity, and the strength of local interactions. To ensure the framework's practicality, we further introduce a measurable proxy for state complexity, thereby linking all core physical quantities in the equation to experimental observables.
    \item Universal Performance Benchmark: Based on this framework, we define a "quantum dynamical performance frontier," constituted by physical processes that achieve maximal efficiency. It provides a clear and universal theoretical standard for evaluating and benchmarking the resource efficiency of any entanglement generation protocol in overcoming locality to create non-local correlations.
    \item Actionable Diagnostic Tool: The ratio of our proposed efficiency factors ($\eta_{\mathrm{LR}} / \eta_{\mathrm{QSL}}$) serves as a powerful diagnostic tool to determine whether the performance bottleneck of a specific process stems from the inefficient utilization of "internal" dynamical resources (low $\eta_{\mathrm{QSL}}$) or from the impeded propagation of "external" correlations within the local system (low $\eta_{\mathrm{LR}}$).
\end{enumerate}

In summary, this theoretical framework opens a new avenue for analyzing and designing complex quantum tasks from the fundamental level of dynamical resource efficiency. The structure of this paper is as follows: Sections \ref{2} and \ref{3} jointly establish the mathematical foundation of our theory. The former rigorously defines the core physical quantities of dynamical resources, information output, and computational complexity, while the latter systematically expounds upon the two performance bounds that serve as our theoretical axioms—the Quantum Speed Limit and the Lieb-Robinson bounds. Building on this foundation, Section \ref{4} rigorously derives our main result—the "energy-entanglement performance equation"—by introducing the core tool of "dynamical efficiency factors." Subsequent sections are dedicated to the examination and interpretation of this core equation: Section \ref{5} proposes a detailed experimental blueprint aimed at testing the fundamental theoretical bounds and characterizing their performance; Section \ref{6} delves into the rich physical implications of the equation, from its application as a diagnostic tool and its profound scale-invariance, to its connections with cutting-edge fields of physics such as complexity theory. Finally, Section \ref{7} summarizes the paper and provides an outlook for future work. To ensure theoretical rigor, a series of key technical details and rigorous proofs are placed in the appendices. These include: complete mathematical derivations for the two fundamental lemmas (Appendices \ref{D} and \ref{E}); a systematic analysis of the mathematical properties of the "dynamical efficiency factors" (Appendix \ref{F}); supplementary clarifications on the $\gamma$ constant, entanglement measures, and core assumptions (Appendices \ref{A}, \ref{B}, and \ref{C}); and Appendix \ref{G}, which provides the complete theoretical foundation and rigorous mathematical proofs for the measurable complexity proxy we introduce. This comprehensive set of appendices provides exhaustive technical support for all the core assertions made in the main text.

\section{Theoretical Framework: System Setup and Core Metrics}
\label{2}
The starting point of our theoretical framework is the objective to establish a quantitative equation that precisely describes the relationship between physical resource consumption and information‑theoretic output in any quantum process. To this end, we consider a lattice system composed of a finite number of qubits (or any finite‑dimensional quantum systems). Its dynamics are governed by a potentially time‑dependent Hamiltonian $H(t)$ that is dominated by local interactions. Specifically, this Hamiltonian can be decomposed into a sum of local terms, $H(t) = \sum_k h_k(t)$, where each term $h_k$ has its support limited to a finite subregion of space. We further assume that the strength of these local interactions is uniformly bounded, i.e., there exists a constant $J$ such that for all terms $h_k$ and all times $t$, their operator norm satisfies $||h_k(t)|| \le J$. This setting is broadly applicable to many important models in condensed matter physics and quantum information science. The physical process of interest is a closed system dynamics that starts from an initial pure state $|\psi(0)\rangle$, undergoes a unitary evolution $U(\Delta t)$ over a time interval $\Delta t$, and reaches a target final state $|\psi(\Delta t)\rangle$. To construct a universal performance equation, we must first rigorously define and quantify the process’s input (dynamical resources), output (entanglement), and intrinsic difficulty (complexity):\cite{Lloyd2000,Nielsen2006,BravyiHastingsVerstraete2006}

\textbf{Definition 2.1 (Available Energy Variance, $\sigma_{\mathrm{avail}}$):} The available energy variance of the system, $\sigma_{\mathrm{avail}}$, is defined as the time-average of the energy standard deviation (or energy fluctuation), $\Delta H(t)$, of the system's Hamiltonian $H(t)$ over the entire evolution interval $[0, \Delta t]$:

\begin{equation}
\sigma_{\mathrm{avail}} = \frac{1}{\Delta t} \int_{0}^{\Delta t} \Delta H(t) \mathrm{d}t = \frac{1}{\Delta t} \int_{0}^{\Delta t} \sqrt{\langle\psi(t)|H(t)^2|\psi(t)\rangle - (\langle\psi(t)|H(t)|\psi(t)\rangle)^2} \mathrm{d}t
\end{equation}
This definition aims to capture the true dynamical resources available to the system for performing any non-trivial quantum operation. Its physical meaning is rooted in the Mandelstam-Tamm formulation of the quantum speed limit: the fundamental "fuel" driving the evolution of a quantum state is not the system's average energy above its ground state, but rather the fluctuation or uncertainty in the energy operator. Only when the system is not in an energy eigenstate (i.e., $\Delta H(t) > 0$) can it undergo non-trivial dynamics under the Hamiltonian evolution. Therefore, $\sigma_{\mathrm{avail}}$ precisely quantifies the average "dynamical activity" available to drive the state's evolution throughout the process.

\textbf{Definition 2.2 (Target Entanglement, $S_E$):} By partitioning the entire system into two complementary subsystems, A and B, we use the entanglement entropy, $S_E$, to measure the correlation between these two subsystems, which represents the core information-theoretic output of the process. In the theoretical derivations of this paper, $S_E$ can generically refer to any valid measure of entanglement. However, to ensure the experimental verifiability of the final theory, the second-order Rényi entropy, $S_E^{(2)} = -\log[\mathrm{Tr}(\rho_A^2)]$, offers a practically more advantageous choice due to its relatively efficient measurement scheme, as detailed in Appendix \ref{B}. Since the core logic of our subsequent derivations and the final equation are universal with respect to the specific choice of entanglement measure, we use $S_E$ for general expression without loss of generality.

\textbf{Definition 2.3 (Optimal Computational Complexity, $C_{\mathrm{opt}}$):} The optimal computational complexity $C_{\mathrm{opt}}$ of a unitary evolution $U(\Delta t)$ is defined as the minimum logical depth among all quantum circuits, composed of some universal gate set, that can realize the target evolution. Here, logical depth refers to the number of sequential operational steps in a quantum circuit that cannot be reduced through parallelization and must be executed in order. Therefore, $C_{\mathrm{opt}}$ is a strictly information-theoretic quantity that quantifies the intrinsic "workload" required to complete the quantum task and sets an insurmountable lower bound on the timescale required for any dynamical process. (The relationship between this theoretical optimum and the achievable circuit complexity in experiments, along with its treatment within this framework, is detailed in Appendix \ref{C.3}.)

\section{Theoretical Foundation: The Two Core Lemmas}
\label{3}
At the core of our theoretical framework is the unification of two seemingly independent universal principles in quantum dynamics. Each of these principles imposes an independent, insurmountable lower bound on the duration$\Delta t$ of a quantum evolution, but from distinct physical perspectives: one rooted in energy uncertainty, the other in the locality of interactions. The purpose of this section is to systematically articulate these two core lemmas, which serve as the axiomatic basis for our subsequent derivations, thereby paving the way for the construction of our final performance equation.\cite{MandelstamTamm1945,MargolusLevitin1998,LiebRobinson1972}The first lemma originates from a Mandelstam-Tamm (MT) type quantum speed limit, which is universally valid for time-dependent systems and profoundly reveals that a quantum system's rate of evolution is constrained by its energy uncertainty. However, to extend this fundamental limit to a complex task possessing an intrinsic computational complexity $C_{\mathrm{opt}}$, we must build a bridge between information theory and physical dynamics. To this end, we posit a core physical assumption: we model the physical realization of a quantum task with optimal complexity $C_{\mathrm{opt}}$ as an idealized process composed of $C_{\mathrm{opt}}$ non-parallelizable, information-theoretically distinguishable elementary steps. Each of these elementary steps physically corresponds to an operation that evolves the system state to an orthogonal one. (This physical assumption is further discussed in Appendix \ref{D}. It therefore serves as an effective model for establishing a lower bound on the dynamical time, based on the task's logical depth.) This assumption translates the abstract "logical depth" into a measurable "dynamical path length" in Hilbert space, thus allowing us to apply the MT bound to set a fundamental lower bound on the "time cost" for completing the entire complex task. This logical chain is formalized as our first core lemma:

\textbf{Lemma 3.1 (MT-type QSL-Complexity Lemma):} Consider a quantum system described by a local Hamiltonian $H(t)$, evolving unitarily from an initial pure state $|\psi(0)\rangle$ over a time $\Delta t$. Let $\sigma_{\mathrm{avail}}$ be the available energy variance of the system during the evolution (as defined in Sec.~\ref{2}), and let $C_{\mathrm{opt}}$ be the optimal computational complexity for realizing this evolution (as defined in Sec.~\ref{2}). Then, the total evolution time $\Delta t$ is bounded by the following inequality\footnote{This lemma is based on the Mandelstam-Tamm bound, which holds rigorously for time-dependent systems. It generalizes the time lower bound for a single-step orthogonal evolution to a complex task composed of $C_{\mathrm{opt}}$ serial logical steps. Its complete and rigorous mathematical derivation is detailed in Appendix \ref{D}.}:

\begin{equation}
\sigma_{\mathrm{avail}} \Delta t \ge \frac{\pi\hbar}{2} C_{\mathrm{opt}} \label{eq:lemma1}
\end{equation}

This "MT-type QSL-Complexity Lemma" profoundly reveals that the duration $\Delta t$ of any quantum process possesses a lower bound determined jointly by its intrinsic computational complexity $C_{\mathrm{opt}}$ and the available energy variance $\sigma_{\mathrm{avail}}$. To accomplish a computationally more complex task in a shorter time, the system must necessarily possess a larger energy fluctuation.
The second lemma originates from the locality of physical interactions. In a system governed by a local Hamiltonian, the propagation of information and the establishment of correlations are not instantaneous processes but are constrained by a finite velocity determined by the local interaction strength. This physical idea is precisely captured by the Lieb-Robinson (LR) bounds. A direct and profound corollary is that the growth rate of quantum entanglement, as a form of non-local correlation, is also subject to a strict upper bound in such systems. Based on this, we can establish our second core lemma:

\textbf{Lemma 3.2 (Locality-Entanglement Lemma):} Consider a quantum many-body system governed by local interactions (with strength upper-bounded by $J$). The system is partitioned into two complementary subsystems, A and B, and their correlation is measured by the entanglement entropy $S_E$. If the system evolves from an initial product state with no entanglement between A and B (i.e., $S_E(0)=0$), then the amount of entanglement $S_E$ that can be generated within a time $\Delta t$ is upper-bounded as follows:

\begin{equation}
S_E \le \gamma \frac{J}{\hbar} \Delta t \label{eq:lemma2}
\end{equation}

where $\gamma$ is a dimensionless constant of order $O(1)$. (A more in-depth discussion of the physical origin of $\gamma$ and its dependence on system geometry is provided in Appendix \ref{A}, while the complete mathematical derivation of this lemma from the Lieb-Robinson bounds is detailed in Appendix \ref{E}.) This "Locality-Entanglement Lemma" reveals another fundamental cost of entanglement generation: time. It asserts that in a world governed by the principle of locality, generating a substantial amount of quantum entanglement necessarily requires a sufficiently long time, the minimum of which is directly constrained by the system's microscopic interaction strength.
These two lemmas jointly delineate the performance boundaries for the dynamical evolution of any local quantum system and serve as the logical starting point for building our unified theory. \textit{(Note: It is crucial to emphasize that the linear growth upper bound for entanglement described in Lemma 3.2 primarily applies to the initial stage of evolution from a product state. For longer evolution times $\Delta t$, due to the finite dimensionality of the Hilbert space, the growth of entanglement entropy $S_E$ will necessarily slow down and eventually saturate. Within our performance equation framework, this saturation effect is perfectly captured by the LR efficiency factor $\eta_{\mathrm{LR}}$, which will be introduced in Sec.~\ref{4}. When entanglement growth enters the saturation regime, the growth rate of $S_E$ is much lower than $J/\hbar$, which, according to the definition of $\eta_{\mathrm{LR}}$ (Eq.~\eqref{eq:4}), will cause $\eta_{\mathrm{LR}}$ to decrease significantly with increasing $\Delta t$. Therefore, our theoretical framework is self-consistent and capable of describing the entire evolution process: in the short-time linear growth regime, $\eta_{\mathrm{LR}}$ approaches its maximum value; in the long-time saturation regime, $\eta_{\mathrm{LR}}$ tends to zero, correctly reflecting the exhaustion of the system's ability to generate new entanglement. The core application scenarios of this paper focus on the performance at the moment the target complexity $C_{\mathrm{opt}}$ and entanglement $S_E$ are reached, and $\eta_{\mathrm{LR}}$ precisely quantifies the overall efficiency of the process along the 'entanglement propagation' pathway up to that point.)}

\section{Main Theorem: The Energy-Entanglement Performance Equation}
\label{4}
The two major lemmas established in the preceding section set independent performance bounds on the duration $\Delta t$ of the same quantum evolution—one originating from energy fluctuation and computational complexity (Lemma \ref{eq:lemma1}), the other from locality and entanglement generation (Lemma \ref{eq:lemma2}). However, a simple conjunction of these two inequalities fails to yield a precise physical law. The fundamental reason is that any real physical process is highly unlikely to saturate both theoretical limits simultaneously. The specific dynamical characteristics of a process determine its actual efficiency in utilizing energy fluctuations and generating entanglement. To establish a truly universal and exact theory, we must transcend this boundary-based analytical paradigm. Our core methodology lies in introducing a pair of dimensionless dynamical efficiency factors, which precisely quantify the proximity of any specific physical evolution to its theoretical performance limits.
First, we define the QSL efficiency factor, $\eta_{\mathrm{QSL}}$:

\noindent\textbf{Definition 4.1 (QSL Efficiency Factor, $\boldsymbol{\eta_{\mathrm{QSL}}}$):} For a quantum process satisfying the conditions of Lemma \ref{eq:lemma1}, its QSL efficiency factor, $\eta_{\mathrm{QSL}}$, is defined as the ratio of the theoretical minimum time cost set by the quantum speed limit to the actual "energy variance-time" cost of the process. Its mathematical form is:

\begin{equation}
\eta_{\mathrm{QSL}} = \frac{(\pi\hbar / 2) C_{\mathrm{opt}}}{\sigma_{\mathrm{avail}} \Delta t}
\label{eq:3}
\end{equation}

According to Lemma \ref{eq:lemma1}, the value of $\eta_{\mathrm{QSL}}$ is always in the range $(0, 1]$. Next, we define the LR efficiency factor, $\eta_{\mathrm{LR}}$.

\noindent\textbf{Definition 4.2 (LR Efficiency Factor, $\boldsymbol{\eta}_{\mathrm{LR}}$):} For a quantum process satisfying the conditions of Lemma \ref{eq:lemma2}, its LR efficiency factor, $\eta_{\mathrm{LR}}$, is defined as the ratio of the entanglement actually generated by the process to the theoretical maximum entanglement that could be generated, as set by the Lieb-Robinson bound. Its mathematical form is:

\begin{equation}
\eta_{\mathrm{LR}} = \frac{S_E}{(\gamma J / \hbar) \Delta t}
\label{eq:4}
\end{equation}

According to Lemma \ref{eq:lemma2}, the value of $\eta_{\mathrm{LR}}$ is always in the range $[0, 1]$. (These efficiency factors are not mere notations but are well-behaved physical quantities, whose rigorous mathematical bounds, analytic behavior, and theoretical calibratability are analyzed in detail in Appendix \ref{F}.)
By introducing these two efficiency factors, we have transformed the two fundamental inequalities into two equalities that hold exactly for any physical process. The evolution time $\Delta t$ is the same physical quantity in both equations, which allows us to eliminate it algebraically, thereby revealing a universal constraint independent of the specific process duration. Solving for $\Delta t$ from Eq.~\eqref{eq:3} and Eq.~\eqref{eq:4} respectively and setting them equal, we have:

\begin{equation}
\frac{(\pi\hbar / 2) C_{\mathrm{opt}}}{\sigma_{\mathrm{avail}} \eta_{\mathrm{QSL}}} = \frac{S_E \hbar}{\eta_{\mathrm{LR}} \gamma J}
\end{equation}

We proceed with an algebraic rearrangement of this expression. First, we note that Planck's constant, $\hbar$, cancels from both sides of the equation. This is not a mere mathematical coincidence but reveals a profound underlying duality. Its necessity is rooted in the fact that the two physical principles forming the foundation of our theory—the QSL and LR bounds—have their respective timescales calibrated by Planck's constant in an identical manner (linear proportionality), as detailed in the derivations in Appendix \ref{D} and Appendix \ref{E}. Consequently, $\hbar$ is inevitably eliminated when the two timescales are equated. After canceling $\hbar$, the equation becomes:

\begin{equation}
\frac{\pi C_{\mathrm{opt}}}{2 \sigma_{\mathrm{avail}} \eta_{\mathrm{QSL}}} = \frac{S_E}{\eta_{\mathrm{LR}} \gamma J}
\end{equation}

Subsequently, by rearranging the above expression to solve for the "resource-output product" ($\sigma_{\mathrm{avail}} S_E$), we arrive at the central result of this paper, which we state as follows:

\noindent\textbf{Theorem 4.1 (Energy-Entanglement Performance Equation, RECT‑$\eta$):} Consider a physical process in a local quantum many-body system designed to generate non-local entanglement. Let $\sigma_{\mathrm{avail}}, S_E, C_{\mathrm{opt}}, J,$ and $\gamma$ be, respectively, the available energy variance, the generated entanglement, the optimal computational complexity, the upper bound on local interaction strength, and the entanglement growth constant for the process. Further, let $\eta_{\mathrm{QSL}}$ and $\eta_{\mathrm{LR}}$ be the efficiency factors for the process, calculated according to Eq.~\eqref{eq:3} and Eq.~\eqref{eq:4}. Then, these physical quantities must satisfy the following exact identity:\footnote{A dimensional analysis and consistency check for all physical quantities in this equation is provided in Appendix \ref{C.1}.}

\begin{equation}
\sigma_{\mathrm{avail}} S_E = \left(\frac{\eta_{\mathrm{LR}}}{\eta_{\mathrm{QSL}}}\right) \left(\frac{\pi\gamma J}{2}\right) C_{\mathrm{opt}}
\label{eq:5}
\end{equation}

\textit{(Note: The dimensionless geometric parameter $\gamma$ is proportional to the size of the bipartition boundary and thus generally depends on the total system size. Its detailed definition and scaling behavior are analyzed in Appendix \ref{A}.)}

This equality, which we term the \textbf{"Energy-Entanglement Performance Equation" (RECT‑$\eta$)}, is the central thesis of this paper. It establishes a universal, efficiency-inclusive performance equation for the physical act of "generating non-local entanglement" at the fundamental level of quantum dynamics. (Note: A rigorous argument for this equation's algebraic nature as a "resource accounting identity" and why its physical content is embodied in the efficiency upper bounds ($\eta \le 1$) can be found in Appendix \ref{C.4}.) The physical implications of this equation are profound and multifaceted:

\begin{enumerate}
    \item \textbf{Resourceful Nature of Entanglement:} This resource identity elevates quantum entanglement from a purely kinematic property to a physical resource whose generation adheres to a precise performance trade-off. Generating any non-zero entanglement entropy $S_E$ necessarily consumes a non-zero budget of "energy variance-complexity" ($\sigma_{\mathrm{avail}} C_{\mathrm{opt}}$).

    \item \textbf{Quantification of Cost:} The equation precisely quantifies this trade-off. A physical process's \textbf{"resource-output product" ($\sigma_{\mathrm{avail}} S_E$)} is quantitatively linked to the task's intrinsic, information-theoretic computational complexity ($C_{\mathrm{opt}}$) via a proportionality factor determined by dynamical efficiencies. This provides a concrete, computable measure for the "cost of non-locality."

    \item \textbf{Physical Composition of the Proportionality Factor:} The proportionality factor in this relationship is determined by two distinct physical components:
    \begin{enumerate}
        \item A fundamental constant ($\pi\gamma J / 2$) determined by the system's intrinsic properties. It is composed of the upper bound on local interaction strength, $J$, and the geometric constant of entanglement growth, $\gamma$. This can be viewed as the base "energy variance-entanglement" cost required per unit of complexity.
        \item A dimensionless efficiency ratio ($\eta_{\mathrm{LR}} / \eta_{\mathrm{QSL}}$) determined by the process's dynamics. This ratio is a novel physical quantity that serves as a diagnostic tool, precisely quantifying whether a process is inefficient in its utilization of energy fluctuations (low $\eta_{\mathrm{QSL}}$) or in the propagation pathway for entanglement generation (low $\eta_{\mathrm{LR}}$).
    \end{enumerate}

    \item \textbf{Universal Calibration and an Intrinsic Duality of Dynamics:} In the final "Energy-Entanglement Performance Equation," Planck's constant $\hbar$ is completely eliminated. This phenomenon reveals an intrinsic duality in the dynamical evolution of local quantum theory. To understand this, we must recognize that our framework essentially performs cost accounting for the same quantum evolution from two fundamentally different yet complementary perspectives, with $\hbar$ acting as a universal calibrator.
    The first perspective is that of evolution in Hilbert space, governed by the Quantum Speed Limit (QSL). It focuses on the motion of the quantum state vector $|\psi(t)\rangle$ within its abstract Hilbert space. The "fuel" for this motion is the energy uncertainty $\sigma_{\mathrm{avail}}$, and its "product" is the information-theoretic computational depth $C_{\mathrm{opt}}$. The role of $\hbar$ here is to serve as the cornerstone of the energy-time uncertainty principle, converting energy (a resource) into a maximal "clock rate" for evolution ($\Delta t \ge \pi\hbar / 2\sigma_{\mathrm{avail}}$). This is purely a description of how a quantum state's internal properties change with time.
    The second perspective is that of propagation in real space, governed by the Lieb-Robinson (LR) bounds. It focuses on how information and correlations propagate across a physical lattice defined by the local interaction strength $J$. Here, the "product" is the non-local entanglement $S_E$ spanning spatial regions. The role of $\hbar$ is to set a fundamental "light cone" or velocity limit for this propagation by calibrating the local interaction strength $J$ ($S_E/\Delta t \le \gamma J/\hbar$). This is purely a description of how the system builds structure in external physical space.
    The physical meaning of our resource identity is that it forms a bridge between these two seemingly independent perspectives. It asserts that in a self-consistent quantum theory evolving under a local Hamiltonian, the total "resource-complexity" cost calculated from the "internal dynamics" perspective is proportional to the total "entanglement generation" cost calculated from the "external correlation" perspective, with the proportionality factor determined by the process efficiencies ($\eta_{\mathrm{LR}} / \eta_{\mathrm{QSL}}$). The disappearance of $\hbar$ reveals the very structure of this bridge. As the universal calibrator that converts energy units (Joules) to frequency units (Hertz), $\hbar$ enters the timescales of both perspectives in an identical (linearly proportional) manner. When we directly compare these two "cost ledgers," which have been gauged by the same ruler, the universal ruler unit itself is naturally cancelled out. The final equation thus emerges as a structural conservation relation governed purely by the system's physical quantities ($\sigma_{\mathrm{avail}}, S_E, J, C_{\mathrm{opt}}$) and the process efficiency factors. This does not imply that the physical process is "non-quantum"; on the contrary, it reveals a profound self-consistency requirement of quantum mechanics itself: the "complexity cost" of a quantum process evolving in Hilbert space is intrinsically dual and mutually constrained with its "structural cost" of building non-local correlations in real space. The form of this constraint relation itself is independent of $\hbar$, the fundamental scale we use to measure "quantumness." This prompts us to elevate the generation of entanglement from a mere corollary of quantum mechanics to a more fundamental physical construction process that follows a universal, quasi-geometric cost law.
\end{enumerate}

Thus far, the "Energy-Entanglement Performance Equation" (Eq.~\eqref{eq:5}) has provided us with a theoretically complete resource identity for accounting for the cost of generating non-locality. The equation reveals a universal trade-off among the system's dynamical resources, information-theoretic output, and intrinsic computational difficulty. However, the direct experimental verification of this framework faces a fundamental challenge: one of its core physical quantities, the optimal computational complexity $C_{\mathrm{opt}}$, is a theoretical limit from information theory that cannot be determined by any known direct physical measurement. This fact presents a practical obstacle to the precise calculation of $\eta_{\mathrm{QSL}}$ and the decisive testing of the entire framework.
To bridge this critical gap between theory and practice, we now introduce a complexity proxy constructed entirely from experimentally measurable quantities, thereby placing the entire theoretical framework on a falsifiable, solid experimental foundation. The physical motivation for our theoretical extension stems from a physical intuition based on causal inference: the very existence of an information-theoretically complex final state (the result) constitutes strong evidence that its generation process must have been complex. We therefore assert that the observed descriptive complexity of the final state can set a strict, non-trivial lower bound on the intrinsic computational complexity of the quantum process that generated it. It is crucial to emphasize that this inferential relationship is unidirectional: while a complex final state necessitates a complex generation process, a complex process does not necessarily produce a complex final state (e.g., when the evolution operator is close to the identity). Our framework is designed to precisely quantify the former case: retroactively determining the minimum complexity cost required to generate an observed, non-trivial final state. (A deeper discussion and formalization are presented in Appendix \ref{G}.) Based on this physical insight, we introduce a "state complexity proxy" that can be efficiently estimated using advanced quantum measurement techniques:

\textbf{Definition 4.3 (State Complexity Proxy, $\widehat{K}_{\mathrm{KL}}$):} Consider a quantum system that evolves from a simple initial state to a final state $\rho_f$. By applying a fixed measurement scheme $M$ (e.g., projective measurements in the computational basis) and performing $N$ independent experimental runs, we obtain an empirical frequency distribution $\{n_i\}$ over a set of $M_{\mathrm{out}}$ distinct measurement outcomes $\{x_i\}$. Its (smoothed) probability distribution is estimated as $\widehat{P}(x_i) = (n_i + \alpha) / (N + M_{\mathrm{out}} \alpha)$. The state complexity proxy of this final state, $\widehat{K}_{\mathrm{KL}}$, is then defined as:

\begin{equation}
\widehat{K}_{\mathrm{KL}} = \log_2(M_{\mathrm{out}}) - D_{\mathrm{KL}}(\widehat{P} || U_{M_{\mathrm{out}}})
\end{equation}

\textit{(Note: This proxy, $\widehat{K}_{\mathrm{KL}}$, is a Shannon quantity calculated from the classical probability distribution $P$ obtained after measurement. Its theoretical value is the Shannon entropy $H(P)$. This concept must be strictly distinguished from the von Neumann entropy $S(\rho)$ of the quantum state $\rho$ itself. Its core role in this framework is to serve as an experimentally accessible proxy that provides a rigorous lower bound for the incomputable algorithmic complexity.On the other hand, before presenting the process–state complexity bound, it is necessary to clarify a crucial prerequisite under which $K_{\mathrm{KL}}$ can serve as an effective experimental proxy for $C_{\mathrm{opt}}$. This hinges on a key physical assumption: the output probability distribution $P$ of a sufficiently deep quantum evolution is algorithmically incompressible, in the sense that its Kolmogorov complexity satisfies $K(P) \gtrsim H(P)$. Intuitively, long-time evolution erases simple structures in the initial state, resulting in an output distribution that appears approximately "pseudorandom" or quantum chaotic; for such distributions, the amount of information required to describe them is on the same order as their Shannon entropy. This assumption is supported not only theoretically---via the typicality arguments of Li and Vitányi (see Appendix~\ref{G})\cite{li2008introduction}---but also empirically, as demonstrated in the randomized circuit sampling experiment by Google Sycamore and the 2025 ion-trap experiment by DeCross et al.\cite{arute2019quantum,liu2025certified} Consequently, as long as the final-state distribution passes standard statistical tests of randomness, $K_{\mathrm{KL}}$ may be regarded as a reliable lower bound on $C_{\mathrm{opt}}$.)}

Here, $D_{\mathrm{KL}}$ is the Kullback-Leibler divergence, and $U_{M_{\mathrm{out}}}$ is the uniform distribution over the $M_{\mathrm{out}}$ outcomes. We emphasize that this proxy can be efficiently and reliably estimated using measurement protocols that are scalable with system size, such as quantum shadow tomography. This state complexity proxy is connected to the logical complexity of the process via the following theorem, which thereby constitutes the bridge connecting our theoretical framework to experimental reality:

\textbf{Theorem 4.2 (Process‑State Complexity Bound):}%
For a unitary evolution $U$ driven by a universal gate set~$G$, starting from a simple initial state with Kolmogorov complexity $O(1)$, its optimal computational complexity $C_{\mathrm{opt}}$ is bounded by the state complexity proxy $\widehat{K}_{\mathrm{KL}}$ of its final state (under measurement scheme $M$):

\begin{equation}
C_{\mathrm{opt}} \;\ge\; c_G\; \bigl(\widehat{K}_{\mathrm{KL}} - K(M)\bigr)\;-\; O\!\bigl(\log\widehat{K}_{\mathrm{KL}}\bigr)
\label{eq:7}
\end{equation}

\textit{(Note: This is the form in the high‑precision limit. To maintain conciseness in the main text, Eq.~\eqref{eq:7} omits the statistical error term originating from finite sampling. This error term vanishes in the high‑precision sampling limit ($\varepsilon\to0$). The full statistical form of this bound and its derivation are detailed in Appendices~\ref{G.2} and~\ref{G.3}.)}%
\cite{MoraBriegel2004,Kaltchenko2021}

Here, $c_G = 1/\log_2|G|$ is a dimensionless "compilation efficiency" constant that depends only on the universal gate set $G$ used, and $K(M)$ is the Kolmogorov complexity of describing the measurement scheme $M$ itself (a small constant that can be calibrated once\footnote{For commonly used measurement schemes like local Clifford shadow tomography \cite{huang2020predicting}, the descriptive complexity of the scheme itself scales linearly with the system size $n$, i.e., $K(M) = O(n)$.}). The rigorous proof of this theorem, which profoundly combines the Solovay-Kitaev theorem with principles from algorithmic information theory, is formally derived and detailed in Appendix \ref{G}.
Having obtained this measurable lower bound, we can now reformulate the QSL inequality, one of the cornerstones of our theory. By substituting this lower bound (Eq.~\eqref{eq:7}) into Lemma \ref{eq:lemma1}, we obtain a new performance boundary built entirely from experimentally measurable quantities. Based on this, we define a new, directly computable proxy QSL efficiency factor, $\eta_{\mathrm{QSL-K}}$:

\begin{equation}
\eta_{\mathrm{QSL-K}} = \frac{(\pi\hbar c_G / 2) (\widehat{K}_{\mathrm{KL}} - K(M))}{\sigma_{\mathrm{avail}} \Delta t} \le 1
\label{eq:8}
\end{equation}

This new efficiency factor $\eta_{\mathrm{QSL-K}}$ corresponds perfectly in physical meaning to $\eta_{\mathrm{QSL}}$, but its calculation no longer depends on the unmeasurable $C_{\mathrm{opt}}$. Finally, by combining it with $\eta_{\mathrm{LR}}$ (whose definition remains unchanged), we arrive at the final, fully experimentally testable Energy-Entanglement Performance Equation:

\begin{equation}
\sigma_{\mathrm{avail}} S_E = \left(\frac{\eta_{\mathrm{LR}}}{\eta_{\mathrm{QSL-K}}}\right) \left(\frac{\pi\gamma J c_G}{2}\right) (\widehat{K}_{\mathrm{KL}} - K(M))
\label{eq:9}
\end{equation}

This final performance equation (Eq.~\eqref{eq:9}) maintains an elegant structural consistency with our initial theoretical result (Eq.~\eqref{eq:5}), but its underlying physical meaning has been deepened. The core complexity term is now represented by a physical quantity, $\widehat{K}_{\mathrm{KL}}$, that can be directly extracted from experimental data, while the proportionality coefficient now incorporates the constant $c_G$, which reflects the hardware's compilation efficiency. This transformation renders a decisive experimental test of our entire theoretical framework possible. It is noteworthy that the efficiency ratio $\eta_{\mathrm{LR}} / \eta_{\mathrm{QSL-K}}$ now becomes an even more powerful diagnostic tool: if its experimentally observed value is significantly greater than 1, it may indicate that the process has reached or surpassed the theoretical performance frontier defined by the current gate set's efficiency. Conversely, a value less than 1 can be used to further diagnose whether the performance bottleneck lies in the gate compilation (either $c_G$ or the circuit structure) or in the energy scheduling ($\sigma_{\mathrm{avail}}$).

\section{Experimental Blueprint: Mapping the Performance Equation and Probing the Frontier}
\label{5}
The ultimate viability of any theory aiming to describe physical reality rests not only on its internal logical consistency but also on its capacity to propose clear, falsifiable physical propositions. Our theoretical framework's core result, the "Energy-Entanglement Performance Equation" (RECT‑$\eta$, Eq.~\eqref{eq:5}), is, by its construction based on the definitions of the efficiency factors, an algebraically exact "resource identity." Therefore, the testable physical content of this experimental framework lies in whether its foundational premises—namely, the efficiency upper bounds $\eta_{\mathrm{QSL}} \le 1$ and $\eta_{\mathrm{LR}} \le 1$ derived from the quantum speed limit and Lieb-Robinson bounds—are respected, and in the framework's effectiveness as a diagnostic and characterization tool. This section details a layered experimental blueprint designed to directly test these fundamental efficiency bounds and, based on this validation, to apply the theoretical framework to the diagnosis of quantum dynamical processes and the benchmarking of their performance frontiers.

\subsection{Experimental Platform and Key Control Parameters}
\label{5.1}
The core of our proposed experimental scheme is to systematically control and precisely measure all key physical quantities involved in the RECT‑$\eta$ equation within a highly controllable quantum system. An ideal experimental platform would be a trapped-ion chain or a superconducting qubit chip comprising tens to hundreds of qubits. These platforms possess the high-fidelity gate operations and high-efficiency quantum state readout capabilities that provide a solid technological foundation for this experiment. The essence of the experimental design lies in having "knobs" for key parameters that can be controlled almost independently. In our scheme, we identify two main classes of core controllable variables:

\begin{enumerate}
    \item \textbf{Dynamical Resource Knob ($\sigma_{\mathrm{avail}}$):} The system's available energy variance, $\sigma_{\mathrm{avail}}$, can be systematically controlled by precisely varying the power or amplitude of the global laser or microwave pulses that drive the quantum evolution. For instance, in a trapped-ion system, one can adjust the intensity of the global laser beam driving the Ising-type interaction; in a superconducting qubit system, one can tune the Rabi frequency $\Omega$ of the driving microwave field. The specific value of $\sigma_{\mathrm{avail}}$ must be determined through detailed calibration of the system's driving Hamiltonian $H(t)$ and precise measurements of its first moment $\langle H(t) \rangle$ and second moment $\langle H(t)^2 \rangle$.
    \item \textbf{Entanglement \& Complexity Knobs ($S_E$ \& $C_{\mathrm{exp}}$):} The target state's entanglement $S_E$ and the applied computational complexity $C_{\mathrm{exp}}$ can be co-regulated by programming and executing quantum circuits of varying structures. For example, starting with a system in an all-aligned product state ($S_E=0$), one can apply a circuit of $D$ layers composed of alternating two-qubit entangling gates (such as CNOT or iSWAP gates). The circuit depth $D$ directly determines the experimentally applied computational complexity, $C_{\mathrm{exp}}$, which constitutes a controllable upper bound on the theoretical optimum, $C_{\mathrm{opt}}$. Following this evolution, the system will reach a final state with a specific entanglement entropy $S_E$. By systematically varying the circuit depth $D$ and the arrangement of gates, a series of final states with different $S_E$ and $C_{\mathrm{exp}}$ can be prepared.
\end{enumerate}
Throughout this experimental framework, we will perform direct and precise measurements of four core dynamical physical quantities: the available energy variance $\sigma_{\mathrm{avail}}$, the final generated entanglement entropy $S_E$ (its specific measurement protocol is detailed in Appendix \ref{B}), the experimentally applied computational complexity $C_{\mathrm{exp}}$, and the total duration of the evolution process $\Delta t$. These observables form the data foundation for all subsequent validation analyses.

\subsection{Experimental Framework: Compliance Check of Fundamental Efficiency Bounds}
\label{5.2}
Before applying our theoretical framework for any diagnosis or characterization, the primary task is to verify that its physical foundations hold within the experimental system under study. This is accomplished by directly testing the two efficiency upper bounds derived from fundamental physical principles (the QSL and LR bounds). We define this process as a two-stage "compliance gate," designed to confirm that the experimental system satisfies the theory's conditions of applicability. Objectivity in this test demands that the background constants in the theoretical model be calibrated through independent, offline methods.

\begin{enumerate}
    \item \textbf{Local Interaction Strength ($J$):} The upper bound on the system's local interaction strength, $J$, can be precisely calibrated through standard spectroscopic analysis or by measuring the Rabi oscillations of a two-body exchange (e.g., iSWAP).
    \item \textbf{Entanglement Growth Constant ($\gamma$):} The dimensionless constant $\gamma$ is closely related to the system's geometric configuration and dimensionality. Its value must be estimated based on the theory in Appendix \ref{A}, combined with a geometric analysis of the system used (e.g., a 1D chain or a 2D lattice) and numerical dynamical simulations.
    \item \textbf{Optimal Computational Complexity ($C_{\mathrm{opt}}$):} $C_{\mathrm{opt}}$ is a theoretical, information-theoretic quantity representing the minimum logical depth required to realize the target unitary evolution. In any specific experiment, $C_{\mathrm{opt}}$ must be estimated via classical, offline computations that are independent of the dynamical evolution. This can be achieved through advanced algorithms such as geometric searches for quantum circuit compilation or methods based on Krylov subspaces. It is crucial that $C_{\mathrm{opt}}$ is not confused with the experimentally applied and typically sub-optimal circuit depth, $C_{\mathrm{exp}}$.
\end{enumerate}
After completing the independent calibrations of $J, \gamma,$ and $C_{\mathrm{opt}}$, we can proceed to a direct test of the two fundamental inequalities that serve as the axioms of our theory.

\begin{enumerate}
    \item \textbf{Gate A (QSL Efficiency Bound Check):} This is a direct test of how the quantum speed limit manifests in complex tasks. The core experimental proposition to be tested is $\eta_{\mathrm{QSL}} \le 1$. Operationally, this means that for each experimental run, the inequality $\sigma_{\mathrm{avail}} \Delta t \ge (\pi\hbar / 2) C_{\mathrm{opt}}$ must be systematically verified by measuring $\sigma_{\mathrm{avail}}$ and $\Delta t$.
    \item \textbf{Gate B (LR Efficiency Bound Check):} This is a direct test of how the principle of locality constrains entanglement generation. The core experimental proposition to be tested is $\eta_{\mathrm{LR}} \le 1$. Operationally, by measuring $S_E$ and $\Delta t$, the inequality $S_E \le (\gamma J / \hbar) \Delta t$ must be systematically verified.
\end{enumerate}
The logic of the test is clear: only when the data from an experimental system consistently and unambiguously pass the checks of both Gate A and Gate B over a wide range of parameters can we consider the system "compliant," i.e., its dynamical behavior falls within the valid descriptive scope of the theoretical framework. The failure of either gate would directly indicate the presence of physical effects beyond the model's assumptions (such as unaccounted-for long-range interactions or strong decoherence noise), which in itself would be an important physical discovery to be reported. This layered verification ensures that any subsequent analysis based on this theoretical framework is built upon a solid and self-consistent physical foundation.

\subsection{Dynamical Diagnosis and Performance Frontier Benchmarking}
\label{5.3}
After confirming the applicability of the theoretical framework through the "compliance checks" of the fundamental laws (Gates A and B), our experimental protocol enters its more constructive application phase. The goal of this phase is to use the theoretical tools we have established—namely, the dynamical efficiency factors—to perform in-depth diagnostics on quantum processes and to experimentally benchmark their ultimate performance boundaries.
One of the most unique contributions of our theoretical framework is the provision of a novel, actionable physical quantity—the efficiency ratio $\eta_{\mathrm{LR}} / \eta_{\mathrm{QSL}}$—for diagnosing the source of a quantum process's performance bottleneck. By experimentally measuring $\sigma_{\mathrm{avail}}, S_E, \Delta t, C_{\mathrm{exp}}$ and substituting them into the defining equations (Eqs.~\eqref{eq:3} and \eqref{eq:4}), we can calculate the efficiency ratio for any specific entanglement generation protocol. This ratio provides a more refined diagnostic insight that goes beyond a mere "total efficiency" assessment:

\begin{enumerate}
    \item When $\eta_{\mathrm{LR}} / \eta_{\mathrm{QSL}} < 1$, it indicates that, relative to the efficiency of utilizing energy fluctuations, the efficiency of entanglement propagation within the system is the more significant performance bottleneck. Optimization strategies should prioritize improving the quantum circuit's structure to enhance $\eta_{\mathrm{LR}}$.
    \item When $\eta_{\mathrm{LR}} / \eta_{\mathrm{QSL}} > 1$, the opposite scenario is revealed: the system's entanglement propagation mechanism is relatively efficient, but the overall utilization of energy fluctuations is insufficient, or the evolution process is too slow. The optimization focus should be on how to drive the system faster or utilize its dynamical resources more efficiently to enhance $\eta_{\mathrm{QSL}}$.
\end{enumerate}
A core task of the experiment is to systematically prepare quantum processes with different dynamical characteristics and measure their efficiency ratios. For instance, we could design and compare two protocols: one being a slow, quasi-adiabatic evolution (expected to have a low $\eta_{\mathrm{QSL}}$) \cite{Farhi2001}, and another being a rapid entanglement circuit utilizing non-local exchange (expected to have a high $\eta_{\mathrm{LR}}$), which can often be realized using techniques like quantum optimal control . Our theory predicts that the measured efficiency ratios will differ significantly in these two cases, thereby validating its predictive power as a diagnostic tool and providing new, first-principles-based guidance for the analysis and optimization of quantum systems.
Another core corollary of our theory is the existence of a "quantum dynamical performance frontier," which represents the optimal trade-off relationship achievable between physical resource investment and information-theoretic output. One of the final goals of the experiment is to map out this frontier with experimental data points in a two-dimensional performance space, with computational complexity $C_{\mathrm{exp}}$ on the x-axis and the "resource-output product" $\sigma_{\mathrm{avail}} S_E$ on the y-axis. The specific procedure is to perform an exploratory scan over a broad parameter space, aiming to find "optimal" dynamical processes that cause the efficiency factors to simultaneously approach their theoretical upper limits (i.e., $\eta_{\mathrm{QSL}} \approx 1$ and $\eta_{\mathrm{LR}} \approx 1$). The experimental data points ($C_{\mathrm{exp}}, \sigma_{\mathrm{avail}} S_E$) corresponding to these optimal processes will form an upper boundary or envelope in this performance space. The shape and slope of this experimentally benchmarked boundary curve directly reflect the ultimate capability of the physical system to efficiently convert abstract computational complexity into a concrete physical resource-output product. It can therefore serve as a universal, first-principles-based benchmark for comparing the intrinsic performance of different physical platforms (e.g., different quantum chip architectures) or different algorithms, providing a new dimension to existing quantum benchmarking methods \cite{eisert2020quantum}.
The applied value of this framework ultimately manifests in its practical output as a benchmarking tool. Once the performance frontier has been experimentally mapped, its upper boundary's behavior can be used to retroactively infer the "compilation efficiency" $r_{\mathrm{exp}} = C_{\mathrm{opt}} / C_{\mathrm{exp}}$. For instance, for points on the performance frontier, we have the approximation $\sigma_{\mathrm{avail}} S_E \approx (\pi\gamma J / 2) C_{\mathrm{opt}}$. By substituting the experimentally measured $\sigma_{\mathrm{avail}} S_E$ and $C_{\mathrm{exp}}$, along with the known $J$ and $\gamma$, an estimate for $r_{\mathrm{exp}}$ can be obtained. This experimentally extracted compilation efficiency provides a quantitative, actionable metric for evaluating and optimizing the implementation efficiency of specific algorithms on particular physical hardware, a challenge of central relevance to the current effort of mapping high-level algorithms to physical gate operations in quantum computing \cite{Farhi2001}.

\subsection{Measurement Protocol for the Complexity Proxy and Closing the Framework Loop}
\label{5.4}
In Sec.~\ref{4}, by introducing the state complexity proxy $\widehat{K}_{\mathrm{KL}}$, we ultimately established a performance equation (Eq.~\eqref{eq:9}) built entirely from experimentally measurable quantities. The final value of this theoretical advance is realized in its testability within a logically clear and technically feasible experimental framework. This section aims to detail such an experimental protocol, whose core lies in the efficient and reliable measurement of $\widehat{K}_{\mathrm{KL}}$, thereby achieving a decisive, closed-loop validation of our entire performance equation framework.
The technical core of our proposed measurement protocol is Quantum Shadow Tomography \cite{huang2020predicting, elben2020mixed}. This technique allows for the efficient extraction of a large amount of classical information from a large-scale quantum state with far fewer measurements than required for full tomography, making it the ideal choice for measuring a many-body observable like $\widehat{K}_{\mathrm{KL}}$. The specific experimental procedure is designed in three main steps:

\begin{enumerate}
    \item \textbf{Step 1 (Shadow Data Acquisition):} On the quantum system that has evolved to the final state $\rho_f$, apply a random unitary transformation drawn from a specific distribution, followed immediately by a global projective measurement in the computational basis. In the most common local Clifford shadow protocol, this step corresponds to independently and uniformly applying a random gate from the single-qubit Clifford group to each qubit of the system. This process (random unitary + projective measurement) is repeated $N_{\mathrm{shot}}$ times, each time starting from the same final state $\rho_f$, and the resulting classical bit strings (the "classical snapshots") as well as the applied random Clifford gates are recorded.

    \item \textbf{Step 2 (Probability Reconstruction):} The $N_{\mathrm{shot}}$ sets of collected data are used for classical post-processing. By applying algorithms such as the median-of-means estimator to all the snapshot data, we can reconstruct a high-fidelity estimate, $\widehat{P}$, of the probability distribution of the final state $\rho_f$ in the computational basis. According to the theory of quantum shadow tomography, for the reconstructed probability distribution $\widehat{P}$ to be within an $l_1$-norm distance of $\varepsilon$ from the true distribution $P$ (i.e., $||\widehat{P} - P||_1 < \varepsilon$) with high confidence $1-\delta$, the required number of measurements $N_{\mathrm{shot}}$ follows the scaling law:
    
    \begin{equation}
    N_{\mathrm{shot}} \ge (c / \varepsilon^2) (\log(M_{\mathrm{out}}) + \log(1/\delta))
    \end{equation}
    
    where $M_{\mathrm{out}}$ is the number of distinct possible measurement outcomes and $c$ is a small numerical constant. This efficient sampling complexity makes it possible to accurately estimate $\widehat{P}$ for medium- to even large-scale quantum systems.

    \item \textbf{Step 3 (Complexity Calculation):} After obtaining the high-fidelity probability distribution estimate $\widehat{P}$ and its outcome space size $M_{\mathrm{out}}$, we can directly calculate the numerical value of the state complexity proxy $\widehat{K}_{\mathrm{KL}}$ using its definition (Definition 4.3).
\end{enumerate}

After completing the efficient measurement of $\widehat{K}_{\mathrm{KL}}$, we must calibrate the compilation efficiency constant $c_G$ in the experimental performance equation (Eq.~\eqref{eq:9}). $c_G$ reflects the intrinsic efficiency of a specific quantum hardware in converting abstract information complexity into physical gate operations. It is not a free fitting parameter but can be calibrated through a one-time benchmarking experiment. A feasible calibration protocol is as follows: select a series of benchmark quantum circuits with known (or tightly bounded) optimal circuit complexities $C_{\mathrm{opt}}$, such as circuits for preparing multi-qubit GHZ or W states. Run these circuits on the experimental device and precisely measure the $\widehat{K}_{\mathrm{KL}}$ of their final states. By performing a linear fit to the experimental data points of $C_{\mathrm{opt}}$ versus $\widehat{K}_{\mathrm{KL}}$, the value of $c_G$ for the hardware platform under the current gate set can be determined. Once this calibration is complete, $c_G$ can be treated as an intrinsic parameter of the device for all subsequent analyses of unknown complex processes.
At this point, the experimental validation framework is fully closed. All physical quantities in Eq.~\eqref{eq:9}—the dynamical resource $\sigma_{\mathrm{avail}}$, the entanglement output $S_E$, the evolution time $\Delta t$, and our newly introduced state complexity proxy $\widehat{K}_{\mathrm{KL}}$—have become experimentally measurable. Experimental researchers are now positioned to carry out a final, decisive verification program consisting of the following three core steps:

\begin{enumerate}
    \item \textbf{Empirical validation of the core physical assumption:} To support the central physical assumption of this theoretical framework—namely, the algorithmic incompressibility of the final-state distribution (see Appendix~\ref{G})—it is recommended that the experimentally reconstructed high-fidelity probability distribution $P$ be subjected to standard classical randomness tests. For example, one may apply the SP 800-22 test suite published by the U.S. National Institute of Standards and Technology (NIST)~\cite{bassham2010sp}. If the distribution $P$ successfully passes these tests, it implies that $P$ does not exhibit easily detectable statistical regularities, thereby providing strong circumstantial evidence in favor of its algorithmic incompressibility and reinforcing the foundation for the subsequent verification steps.
    \item \textbf{Direct Test of the Proxy QSL Bound:} By substituting all measured values into Eq.~\eqref{eq:8}, directly calculate $\eta_{\mathrm{QSL-K}}$ and verify that it strictly satisfies the physical bound of $\le 1$. This would constitute the most direct and powerful test of our framework's foundation.
    \item \textbf{Mapping the Experimental Performance Frontier:} Plot the experimental data points for different dynamical processes with the measured $\widehat{K}_{\mathrm{KL}}$ as the x-axis and the "resource-output product" $\sigma_{\mathrm{avail}} S_E$ as the y-axis. Directly compare these points with the theoretical performance frontier defined by Eq.~\eqref{eq:9}, whose slope is determined by the system's intrinsic parameters ($J, \gamma, c_G$). This not only validates the predictive power of the theory but also provides an unprecedented, first-principles-based quantitative tool for evaluating and benchmarking the resource efficiency of different quantum algorithms or platforms.
\end{enumerate}

\section{Discussion: Physical Implications of the Performance Equation}
\label{6}
The "Energy-Entanglement Performance Equation" (RECT‑$\eta$, Eq.~\eqref{eq:5}) framework we have constructed is, at its core, an algebraically exact resource identity. It fills a gap left by traditional inequality-based paradigms, providing a precise, efficiency-based language for analyzing the process of generating non-local entanglement. The physical implications embodied in this framework are profound and multifaceted. This section aims to systematically explore these implications to elucidate the physical picture underpinning the framework and its connections to the fundamental principles of quantum mechanics. We will first analyze its most direct corollary: the existence of a universal quantum dynamical performance frontier. Subsequently, we will delve into the utility of the newly introduced efficiency ratio as a diagnostic tool for dynamics, followed by an examination of the universal calibration and dynamical duality revealed by the framework. Finally, we will situate our research findings within the broader contemporary frontier of physics concerning complexity theory and the holographic principle.

\subsection{The Quantum Dynamical Performance Frontier}
\label{6.1}
One of the most direct and significant physical corollaries of our theoretical framework arises from the two fundamental efficiency upper bounds upon which it rests ($\eta_{\mathrm{QSL}} \le 1$ and $\eta_{\mathrm{LR}} \le 1$). These two upper bounds naturally define the existence of a \textbf{"Quantum Dynamical Performance Frontier."} This frontier represents the optimal trade-off relationship achievable between the physical resources invested in a quantum process and the information-theoretic output it can produce.
To formalize this concept, we consider the ideal limit where a physical process operates at maximum efficiency. This corresponds to the scenario where both the QSL efficiency factor $\eta_{\mathrm{QSL}}$ and the LR efficiency factor $\eta_{\mathrm{LR}}$ simultaneously approach their theoretical maximum value of 1. In this limit ($\eta_{\mathrm{QSL}} \to 1, \eta_{\mathrm{LR}} \to 1$), our resource identity (Eq.~\eqref{eq:5}) simplifies to a more direct approximate equality:
\begin{equation}
\sigma_{\mathrm{avail}} S_E \approx \left(\frac{\pi\gamma J}{2}\right) C_{\mathrm{opt}}
\end{equation}
This simplified relation defines the performance frontier. It can be intuitively understood as a linear boundary in an abstract performance space with the "resource-output product" ($\sigma_{\mathrm{avail}} S_E$) as one axis and the "optimal computational complexity" ($C_{\mathrm{opt}}$) as the other. The existence of this boundary is a universal assertion: any physical process governed by local interactions must have its state point lie on or below this line. Regardless of its specific dynamical details, no process can achieve a resource-output product that surpasses the limit determined by its intrinsic complexity and the system's fundamental physical parameters.
The primary significance of this frontier is its existence as a \textbf{universal theoretical benchmark}. It provides a clear, quantitative criterion for judging any real-world entanglement generation protocol—be it an algorithm executed on a quantum processor or the natural evolution of a many-body system. The "distance" of a specific process's state point from this frontier can serve as an absolute measure of its overall dynamical efficiency. Furthermore, the slope of this frontier, $k = (\pi\gamma J / 2)$, is determined entirely by the system's intrinsic physical properties: namely, the upper bound on local interaction strength, $J$, and the dimensionless geometric constant related to entanglement propagation, $\gamma$. This point has a crucial practical implication: by experimentally measuring the performance of a series of near-optimal processes and determining their performance boundary, one could, in principle, perform a "dynamical measurement" in reverse to infer the product value of $\gamma J$ for that specific quantum system. This establishes a novel, verifiable bridge between macroscopic performance benchmarking and the microscopic parameters of the underlying Hamiltonian.

\subsection{The Efficiency Ratio as a Dynamical Diagnostic Tool}
\label{6.2}
A core methodological innovation of our theoretical framework is the introduction of the dynamical efficiency factors ($\eta_{\mathrm{QSL}}$ and $\eta_{\mathrm{LR}}$). They not only enable the construction of a self-consistent resource accounting framework but also give rise to a novel, dimensionless observable with profound physical meaning—the efficiency ratio $\eta_{\mathrm{LR}} / \eta_{\mathrm{QSL}}$. This ratio is not a simple mathematical construct but a powerful \textbf{diagnostic tool} capable of revealing the underlying source of a quantum process's performance bottleneck.
To understand its diagnostic value, we recall the intrinsic duality picture proposed in Sec.~\ref{4}: $\eta_{\mathrm{QSL}}$ quantifies the efficiency of utilizing dynamical resources within the system's \textbf{"internal" Hilbert space}, while $\eta_{\mathrm{LR}}$ quantifies the efficiency of propagating entanglement correlations in \textbf{"external" physical space}. Therefore, the efficiency ratio $\eta_{\mathrm{LR}} / \eta_{\mathrm{QSL}}$ directly compares the relative performance of these two independent efficiency channels:
\begin{enumerate}
    \item When $\eta_{\mathrm{LR}} / \eta_{\mathrm{QSL}} < 1$, it implies $\eta_{\mathrm{LR}} < \eta_{\mathrm{QSL}}$. This indicates that, relative to the system's efficiency in utilizing its internal energy fluctuations, its efficiency in generating and propagating entanglement in external space is the more significant performance bottleneck. In other words, the system's energy uncertainty might be efficiently used to drive state evolution, but this evolution does not effectively translate into the growth of non-local correlations between the two subsystems. This could stem from an improper design of the quantum circuit or from structural factors within the system that impede the effective propagation of entanglement.
    \item When $\eta_{\mathrm{LR}} / \eta_{\mathrm{QSL}} > 1$, it implies $\eta_{\mathrm{LR}} > \eta_{\mathrm{QSL}}$. This situation reveals the opposite source of the performance bottleneck. It suggests that the system's entanglement propagation mechanism is relatively efficient, but its overall internal dynamical resource utilization is low. A typical example would be a process that proceeds at a slow pace, far exceeding its theoretical minimum computation time (i.e., very large $\Delta t$), which would cause $\eta_{\mathrm{QSL}}$ to become very small. Although entanglement might be effectively established during this process, the overall efficiency is extremely low due to "idling" in the utilization of dynamical resources.
    \item When $\eta_{\mathrm{LR}} / \eta_{\mathrm{QSL}} \approx 1$, it implies that the system has achieved a degree of \textbf{dynamical balance} between the efficiency of utilizing internal dynamical resources and the efficiency of generating external correlations. The ideal process on the performance frontier is the ultimate manifestation of this balance ($\eta_{\mathrm{QSL}}$ and $\eta_{\mathrm{LR}}$ both approaching 1).
\end{enumerate}
This analysis reveals the diagnostic value of the efficiency ratio. Experimentally, by independently measuring $\sigma_{\mathrm{avail}}, S_E, C_{\mathrm{exp}},$ and $\Delta t$, we can calculate the efficiency ratio for any specific entanglement generation protocol. This ratio provides a more refined diagnostic insight that goes beyond a mere "total efficiency" assessment. For example, suppose two different quantum algorithms exhibit similar overall "energy variance-to-entanglement" conversion efficiency, but one has an efficiency ratio much less than 1, while the other's is close to 1. Our framework suggests that the optimization strategy for the first algorithm should prioritize improving its quantum circuit structure to enhance entanglement propagation (increase $\eta_{\mathrm{LR}}$), whereas optimizing the second algorithm might require a more holistic consideration of how to shorten the total evolution time or more efficiently utilize energy fluctuations. Thus, the efficiency ratio $\eta_{\mathrm{LR}} / \eta_{\mathrm{QSL}}$ transforms from a purely theoretical construct into a physical quantity with clear operational meaning. It provides us with a "prism" that can decompose the overall efficiency exhibited by a complex quantum dynamical process into its two fundamental components, originating from internal dynamical resource utilization and external information propagation, thereby offering new, first-principles-based guidance for the analysis and optimization of quantum systems. This diagnostic capability can be directly verified within the experimental framework proposed in Sec.~\ref{5}.

\subsection{Universal Calibration, Dynamical Duality, and the Uncertainty Principle from a Generative Perspective}
\label{6.3}
The physical substance of the resource identity framework we have constructed is embodied in the complete cancellation of Planck's constant, $\hbar$, in the performance equation (Eq.~\eqref{eq:5}). This phenomenon reveals an \textbf{intrinsic duality} in the dynamical evolution of local quantum theory. To understand this, we must recognize that our framework essentially performs cost accounting for the same quantum evolution from two fundamentally different yet complementary perspectives, with $\hbar$ acting as a \textbf{universal calibrator}.
The first perspective is that of evolution in Hilbert space, governed by the Quantum Speed Limit (QSL). It focuses on the motion of the quantum state vector $|\psi(t)\rangle$ within its abstract Hilbert space. The "fuel" for this motion is the energy uncertainty $\sigma_{\mathrm{avail}}$, and its "product" is the information-theoretic computational depth $C_{\mathrm{opt}}$. The role of $\hbar$ here is to serve as the cornerstone of the energy-time uncertainty principle, converting energy (a resource) into a maximal "clock rate" for evolution ($\Delta t \ge \pi\hbar / 2\sigma_{\mathrm{avail}}$). This is purely a description of how a quantum state's internal properties change with time.
The second perspective is that of propagation in real space, governed by the Lieb-Robinson (LR) bounds. It focuses on how information and correlations propagate across a physical lattice defined by the local interaction strength $J$. Here, the "product" is the non-local entanglement $S_E$ spanning spatial regions. The role of $\hbar$ is to set a fundamental "light cone" or velocity limit for this propagation by calibrating the local interaction strength $J$ ($S_E/\Delta t \le \gamma J/\hbar$). This is purely a description of how the system builds structure in external physical space.
Our resource identity thus forms a bridge between these two seemingly independent perspectives. It asserts that in a self-consistent quantum theory evolving under a local Hamiltonian, the total "resource-complexity" cost calculated from the "internal dynamics" perspective must be proportional to the total "entanglement generation" cost calculated from the "external correlation" perspective, with the proportionality factor determined by the process efficiencies ($\eta_{\mathrm{LR}} / \eta_{\mathrm{QSL}}$). The disappearance of $\hbar$ reveals the very structure of this bridge. As the universal calibrator that converts energy units (Joules) to frequency units (Hertz), $\hbar$ enters the timescales of both perspectives in an identical (linearly proportional) manner. When we directly compare these two "cost ledgers," which have been gauged by the same ruler, the universal ruler unit itself is naturally cancelled out.
The final equation thus emerges as a \textbf{structural conservation relation} governed purely by the system's physical quantities ($\sigma_{\mathrm{avail}}, S_E, J, C_{\mathrm{opt}}$) and the process efficiency factors. This does not imply that the physical process is "non-quantum"; on the contrary, it reveals a profound self-consistency requirement of quantum mechanics itself: the "complexity cost" of a quantum process evolving in Hilbert space is intrinsically dual and mutually constrained with its "structural cost" of building non-local correlations in real space. The form of this constraint relation itself is independent of $\hbar$, the fundamental scale we use to measure "quantumness." This prompts us to elevate the generation of entanglement from a mere corollary of quantum mechanics to a more fundamental physical construction process that follows a universal, quasi-geometric cost law.

This \textbf{"generative perspective"} also offers a new vantage point from which to view the cornerstone of quantum mechanics—the Heisenberg uncertainty principle. Traditionally, the uncertainty principle is often interpreted as a limitation on the act of "measurement." In our framework, however, it is more profoundly embodied as a fundamental law governing the cost that must be paid to "generate" physical reality. For example, the energy-time uncertainty can be understood as: to generate a dynamical process that is highly localized in time (small $\Delta t$), one must pay a "resource budget" that has a large fluctuation in energy (large $\sigma_{\mathrm{avail}}$). Our MT-type QSL-Complexity Lemma directly links this principle to the information-theoretic difficulty of the task, $C_{\mathrm{opt}}$, thereby "pricing" the generation of a process with a specific complexity. Our work does not challenge the uncertainty principle but rather provides a complementary physical picture based on \textbf{"generative dynamics" and "information complexity."} In this picture, quantum uncertainty is not only an intrinsic limit on measurement but also a universal resource trade-off law that must be obeyed when actively constructing physical reality.

\subsection{Connections to Complexity Theory and the Holographic Principle}
\label{6.4}
The significance of the theoretical framework we have constructed is not limited to the direct assessment of resource efficiency in local quantum systems; it lies in its ability to establish new, testable links with some of the core frontier ideas in contemporary fundamental physics. In high-energy physics and quantum gravity research, "complexity" is evolving from an abstract information-theoretic concept into a core physical quantity believed to be dual to fundamental physical properties such as spacetime geometry. Our work can be seen as a \textbf{bottom-up}, operational example of this grand narrative, originating from local quantum many-body systems.
This conceptual convergence is first reflected in its resonance with concepts like "operator complexity," with the $C_{\mathrm{opt}}$ in our theory being in the same spirit. More profoundly, our framework endows these theories with a concrete physical cost and establishes a path from abstract operator complexity to measurable final-state complexity. This particularly resonates deeply with explorations of the physics inside black holes within the \textbf{Holographic Duality (AdS/CFT Correspondence)}. A series of influential conjectures proposed by Susskind and others, such as "Complexity equals Volume/Action" (CV/CA), posits at its core that the complexity of a quantum state in the boundary field theory (CFT) is dual to the geometry of the gravitational bulk (AdS).\cite{Maldacena1999,Susskind2016,Brown2016} However, these profound top-down insights still lack a microscopic definition of complexity from the boundary field theory that can be directly calculated. Our work plays a unique and illuminating complementary role here. Within the framework of a non-gravitational quantum many-body system with a well-defined local Hamiltonian, we have established, through first-principles construction, a quantitative relationship that can be experimentally tested.
Specifically, our final performance equation (Eq.~\eqref{eq:9}) forges a precise, dynamically-efficiency-modulated quantitative bridge among three types of physical quantities: dynamical resources (measured by energy variance $\sigma_{\mathrm{avail}}$), system structure and correlations (measured by non-local entanglement $S_E$), and information-theoretic complexity (measured by the testable state complexity proxy $\widehat{K}_{\mathrm{KL}}$). This "complexity-structure-resource" relationship depicted by Eq.~\eqref{eq:9} offers beneficial support to those grander holographic conjectures on two levels:
\begin{enumerate}
    \item It concretely realizes and validates the core idea that "complexity is a physical quantity associated with the system's macroscopic properties (like entanglement)" in a more fundamental and controllable setting.
    \item It reveals a possible mathematical form for this association, whose independence from Planck's constant $\hbar$ and dependence on process efficiency might offer new structural insights for the future search for more complex observables in boundary field theory that are dual to gravitational geometry.
\end{enumerate}
It must be emphasized that we do not claim to have provided a microscopic proof for any holographic conjecture. Our theoretical framework is aligned in spirit and goal with these frontier conjectures: both are dedicated to understanding how abstract computational complexity manifests itself in the physical world. Our work, as a theoretically and experimentally closed-loop framework, can be seen as a concrete realization within a more fundamental setting—one that is, in principle, testable in a laboratory—under this grand program, thereby opening a new, non-gravitational theoretical and experimental path for exploring the deep connections among information, dynamics, and the structure of physical reality.

\subsection{A Dynamical Inquiry into the Cost of Non-local Correlation}
\label{6.5}
Since the seminal work of Einstein, Podolsky, and Rosen \cite{Einstein1935}, the non-local correlations predicted by quantum mechanics—vividly termed "spooky action at a distance" (spukhafte Fernwirkung)—have been at the heart of speculation and exploration in fundamental physics. Although subsequent theoretical and experimental advances, particularly the formulation of Bell's inequalities \cite{Bell1964} and their experimental verification, have profoundly shaped our understanding of non-locality, the deep question of its physical origin remains of significant theoretical value. Our theoretical framework offers a complementary, resource-based dynamical perspective on this issue.
Our work's starting point is to focus on a slightly different question: in a physical system that strictly adheres to the principle of locality, what is the quantitative physical cost required to generate such non-local correlations? Our theoretical framework provides a possible, computable answer, which may, in turn, offer new clues to understanding the origin of non-locality. Our core performance equation (Eq.~\eqref{eq:9}) reveals that the generation of any non-zero entanglement entropy $S_E$ (as a measure of non-local correlation) is necessarily linked to a series of physical quantities that strictly obey the principle of locality. Specifically, to establish non-local correlations in an initially uncorrelated system, the system must:
\begin{enumerate}
    \item Be injected with a non-zero energy variance $\sigma_{\mathrm{avail}}$ as the source of dynamical driving;
    \item Undergo an evolution for a finite duration $\Delta t$;
    \item Rely on local interactions of strength $J$ to propagate correlations, a process constrained by the causal light cone set by the Lieb-Robinson bounds \cite{BravyiHastingsVerstraete2006}.
\end{enumerate}
This implies that any eventually manifested non-local correlation can be seen as the necessary outcome of a dynamical evolution that is entirely local in time and space and follows the laws of physical causality. From this perspective, when we observe correlations between two spatially separated particles A and B, we might interpret this as a traceable consequence of their shared dynamical history, which is constrained by our performance equation, rather than some instantaneous, mysterious interaction. We do not seek to demystify the profound nature of non-local correlation itself, but rather to propose that its "spooky" character may be rooted in a real, costly, and entirely non-superluminal generation process.
Therefore, our work could be seen as supplementarily extending the discussion on non-locality from a kinematic problem focused on the statistics of measurement outcomes to a resource accounting problem focused on \textbf{"generative dynamics."} Within this framework, the performance equation (Eq.~\eqref{eq:9}) provides a quantitative ledger for this "generation cost." It shows that there is a strict lower bound on the minimum physical resources (energy, time, interaction strength) required to obtain a final state with a specific degree of non-locality. This is in the same spirit as "Quantum Resource Theories" in modern quantum information, which treat entanglement and other non-classical features as valuable resources that need to be "manufactured" through physical operations \cite{Chitambar2019}.
In summary, we hope that this analysis, based on dynamics and resource theory, will provide a useful complementary perspective on the fundamental issue of quantum non-locality. Research indicates that nonlocal quantum phenomena may be underpinned by a generative history that is precisely quantifiable and governed by local physical laws.The full, computable "cost" of the so-called physical "action" may have already been paid for, locally.

\section{Conclusion and Outlook}
\label{7}
In this paper, we have provided a systematic and, ultimately, fully experimentally verifiable theoretical framework for the fundamental question of the "physical cost of generating non-local entanglement under locality constraints." By introducing "dynamical efficiency factors" as a core analytical tool, we have unified the quantum speed limit and the Lieb-Robinson bounds into an "Energy-Entanglement Performance Equation." Crucially, by further introducing a measurable state complexity proxy, we have paved the way for the complete experimental validation of this theoretical framework, transforming it from a theoretical construct into a fully operational physical theory.
The significance of this framework extends beyond the equation itself. It recasts quantum entanglement from an abstract property into a physical resource whose generation follows a precise and directly measurable cost-accounting. This accounting system clearly reveals how a process's "energy variance-entanglement product" is quantitatively linked, via its dynamical efficiencies, to the task's measurable information complexity and the system's local interaction strength. Based on this, our theoretical framework defines a "performance frontier" that can be directly benchmarked experimentally and provides a suite of dynamical tools for diagnosing performance bottlenecks.
The value of any fundamental theoretical exploration lies not only in answering old questions but also in revealing new ones worthy of investigation. Our rigorous theoretical framework naturally delineates its own boundaries of applicability, thereby opening up broad new territories for future research. Among these, three directions are particularly crucial and challenging:

\begin{enumerate}
    \item \textbf{From Local to Long-Range Interactions:} Our theory is built upon the assumption of strictly local interactions. A highly appealing direction is to generalize it to systems with algebraically decaying long-range interactions. In such systems, the form of the Lieb-Robinson bounds itself is altered, and we conjecture that the structure of the performance equation might assume a new, elegant form dependent on the interaction's decay exponent. Overcoming this challenge would not only significantly broaden the universality of our theory but could also provide theoretical guidance for understanding and designing novel quantum materials or quantum networks with non-local connectivity.

    \item \textbf{From Closed to Open Systems:} How to extend this framework to open quantum systems that exchange heat with an environment is a more profound and practically relevant challenge. In a real, dissipative world, the utility of dynamical resources becomes more complex. We anticipate that a future development could establish a quantitative link between our defined "available energy variance" and core concepts in non-equilibrium quantum thermodynamics, such as "non-equilibrium free energy" or "total entropy production rate." Successfully building this bridge would position our performance-equation framework as a cornerstone connecting quantum information, quantum dynamics, and non-equilibrium quantum thermodynamics, and provide a fundamental theoretical tool for assessing and optimizing the ultimate energy-efficiency limits of real-world quantum devices, such as quantum processors and sensors.

    \item \textbf{Measurement and Application of Complexity:} Our introduced measurable complexity proxy, $\widehat{K}_{\mathrm{KL}}$, in and of itself, constitutes a powerful new tool. Future research could focus on systematically measuring the $\widehat{K}_{\mathrm{KL}}$ values for different physical platforms and algorithms, and combining this with the calibration of the compilation efficiency constant $c_G$ to establish a new paradigm for quantum hardware benchmarking. Furthermore, applying this complexity measurement tool to other physical problems, such as diagnosing quantum chaos or characterizing many-body localization, constitutes an extremely appealing avenue of exploration.
\end{enumerate}

Ultimately, our work points toward a more fundamental physical picture: the generation of all structures in the physical reality we perceive, be they local or non-local, is fundamentally governed by universal and quantitatively describable resource trade-off laws. We conclude this paper with a concise assertion: The cost of non-locality is real, computable, and, ultimately, measurable.


\appendix
\section{On the Entanglement Growth Rate Constant \texorpdfstring{$\gamma$}{gamma}}
\label{A}
In the derivation of this paper's main theorem, a key physical input is the "Locality-Entanglement Lemma" (Eq.~\eqref{eq:lemma2}): $S_E \le \gamma \frac{J}{\hbar} \Delta t$. The core of this lemma is that it sets an upper bound on the entanglement growth rate, which is jointly determined by the local interaction strength $J$, Planck's constant $\hbar$, and a dimensionless constant $\gamma$. Although we treated $\gamma$ as a constant of order $O(1)$ in the main text to maintain the theory's generality and simplicity, its specific origin, physical meaning, and behavior in different physical systems are crucial for precise theoretical predictions and experimental design. This appendix aims to provide a more in-depth technical clarification of the constant $\gamma$.\cite{BravyiHastingsVerstraete2006,VanAcoleyen2013}

\subsection{Physical Origin and Dimension of \texorpdfstring{$\gamma$}{gamma}}
\label{A.1}
The constant $\gamma$ originates directly from rigorous mathematical proofs on the upper bounds of entanglement dynamics in quantum many-body systems. Its theoretical foundation is the Lieb-Robinson (LR) bounds, which set an exponentially decaying bound for the "out-of-light-cone" part of operators in systems with local interactions. Based on the LR bounds, pioneering work by Bravyi, Hastings, Verstraete, and others has shown that for a lattice system partitioned into subsystems A and B, the growth rate of the von Neumann entropy $S_E$ between them, $\mathrm{d}S_E/\mathrm{d}t$, is subject to an upper bound proportional to the size of the partition boundary, $|\partial A|$. Its basic form can be written as:
\begin{equation}
\frac{\mathrm{d}S_E}{\mathrm{d}t} \le c \frac{J}{\hbar} |\partial A|
\label{eq:A1}
\end{equation}
Here, $J$ is the upper bound on the local interaction strength we defined (with dimension of energy [E]), and $\hbar$ is the reduced Planck constant (with dimension of energy $\times$ time [E$\cdot$T]), so $J/\hbar$ has the dimension of frequency [1/T]. $|\partial A|$ is the number of sites (or bonds) on the boundary of subsystem A, which is a dimensionless integer. $c$ is a purely dimensionless constant that does not depend on the system size $N$, with its value related to geometric details such as the system's dimension and lattice coordination number.
In the setting of this paper, we consider a bipartition that divides the entire system in two. In this case, the size of the boundary $|\partial A|$ is itself a quantity of order $O(1)$ that depends on the system's geometry. Therefore, to simplify the expression, we absorb the product of these two dimensionless quantities, $c$ and $|\partial A|$, into a new, more general dimensionless constant $\gamma$, defining $\gamma \equiv c \cdot |\partial A|$. With this definition, integrating Eq.~\eqref{eq:A1} with respect to time (assuming evolution starts from $S_E(0)=0$) directly yields Lemma 3.2 used in our main text:
\begin{equation}
S_E \le \gamma \frac{J}{\hbar} \Delta t
\label{eq:A2}
\end{equation}
Thus, the constant $\gamma$ encapsulates the maximum "flux" of entanglement that can "flow" across the partition boundary, as determined by the system's spatial dimension and geometry. It is a purely dimensionless numerical value of order $O(1)$.

\textit{(Note: To ensure a precise understanding of the properties of $\gamma$, we must clarify its scaling behavior. $\gamma$ is a dimensionless parameter uniquely determined by the system's geometry. For a system of a fixed size and geometry, $\gamma$ is a definite constant. However, when comparing systems of different sizes ($L$), the value of $\gamma$ is proportional to the size of the bipartition boundary $|\partial A|$ and thus typically grows with the system size (e.g., in a $d$-dimensional system, for a bipartition, $|\partial A|$ is proportional to $L^{d-1}$). Therefore, calling $\gamma$ 'of order $O(1)$' in the context of this manuscript is intended to emphasize its dimensionless nature and its independence from any dynamical details (such as $\sigma_{\mathrm{avail}}$, $\Delta t$, $C_{\mathrm{opt}}$), rather than implying it is a constant strictly independent of system size. This scaling dependence is an intrinsic feature of the theoretical framework, correctly reflecting the physical fact that larger or higher-dimensional systems have a higher theoretical upper limit for their entanglement generation rate. This also means that in our final performance equation, $\gamma$ plays the role of a core parameter characterizing the 'channel capacity' of a specific system's geometry.)}

\subsection{Behavior of \texorpdfstring{$\gamma$}{gamma} in Different Systems}
\label{A.2}
The precise numerical value of $\gamma$ is crucial for a quantitative test of our performance equation. Its value is not a universal constant of nature but depends on the specific physical system.

\begin{enumerate}
    \item \textbf{1D Systems:} For one-dimensional spin chains with nearest-neighbor interactions, relevant theoretical studies have provided good estimates for $\gamma$. For instance, for the von Neumann entropy, it has been rigorously proven that $c$ is a constant of order $O(1)$. Given that the bipartition boundary $|\partial A|$ for a 1D chain is typically 2, $\gamma$ is also a constant of order $O(1)$. Specific numerical calculations and simulations show that for certain integrable models, the value of $\gamma$ can be accurately estimated.

    \item \textbf{Higher-Dimensional Systems:} In two- or three-dimensional systems, the value of $\gamma$ is significantly influenced by the lattice geometry (e.g., square or honeycomb lattices) and the coordination number, as these factors directly determine the size of the boundary $|\partial A|$. In general, higher dimensions and coordination numbers allow for a larger entanglement flux, leading to an increased value of $\gamma$. This means that in higher-dimensional systems, generating the same amount of entanglement may theoretically require less time (if dynamical resources and complexity permit), but the form of the corresponding energy-entanglement performance equation remains unchanged; only the "base unit price" or the slope of the performance frontier will increase due to the larger value of $\gamma$.

    \item \textbf{Systems with Long-Range Interactions:} The core derivations in this paper rely on the assumption of strict locality. For systems with long-range interactions of the form $1/r^\alpha$, the form of the Lieb-Robinson bounds itself changes. When $\alpha$ is greater than the system dimension $d$, the system still possesses a finite, albeit more complex, upper bound on the propagation speed. In this case, the theoretical framework remains applicable in principle, but the calculation of the constant $\gamma$ will require more refined analysis. However, when $\alpha \le d$, a strict LR bound may no longer hold. In this regime, our foundational "Locality-Entanglement Lemma" may be invalidated, and the applicability of our performance equation would need to be re-examined. This constitutes an important direction for future research.
\end{enumerate}
In summary, the constant $\gamma$ is the key dimensionless parameter that connects microscopic dynamics to macroscopic entanglement generation. It is not only the foundation that determines the upper limit of entanglement growth in our theory but also plays a decisive role in the final performance equation (Eq.~\eqref{eq:5}) by setting the theoretical upper limit of the quantum dynamical performance frontier. Although its specific value depends on system details, this does not affect the universality of our performance equation. Instead, it opens up new possibilities for experimentally measuring the performance frontier and, in turn, precisely inferring the value of $\gamma$ for different quantum systems. In any specific experimental verification, an independent theoretical estimation or numerical simulation of the $\gamma$ value for the platform used will be a critical step to ensure a precise comparison between experimental results and theoretical predictions.

\section{Experimental Protocol for Measuring Target Entanglement \texorpdfstring{$S_E$}{SE}}
\label{B}
The experimental blueprint outlined in Sec.~\ref{5} has at its core the decisive quantitative validation of the "Energy-Entanglement Performance Equation." The cornerstone of this validation is the efficient and precise measurement of the entanglement entropy, $S_E$, generated in the final state of the dynamical evolution. Traditional Quantum State Tomography, due to its measurement cost scaling exponentially with the system size $N$, is practically infeasible for the medium-scale systems ($N > 20$) of interest to us, which are capable of exhibiting complex many-body dynamics. Therefore, we must resort to more advanced and scalable measurement paradigms. This appendix is dedicated to a detailed technical exposition of the protocol we employ.

\subsection{Randomized Measurements and Purity Estimation}
\label{B.1}
The decisive experimental validation of the energy-entanglement performance equation proposed in this paper rests on the efficient and precise measurement of the entanglement entropy, $S_E$, generated in the system's final state. However, traditional Quantum State Tomography is practically infeasible for the medium-scale systems of interest, which can exhibit complex many-body dynamics, as its measurement cost scales exponentially with the subsystem size. Therefore, to ensure the practical feasibility of our experimental blueprint, we must resort to an advanced measurement paradigm that is more scalable in terms of resource consumption.
To this end, we select the second-order Rényi entropy, $S_E^{(2)} = -\log[\mathrm{Tr}(\rho_A^2)]$, as our core entanglement measure. Its fundamental advantage lies in its direct connection to a more experimentally accessible quantity: the purity of the reduced density matrix $\rho_A$, given by $P(\rho_A) = \mathrm{Tr}(\rho_A^2)$. Thus, the challenge of measuring entanglement entropy is rigorously transformed into the task of measuring purity.
The protocol we adopt is based on the framework of Randomized Measurements, pioneered by Elben, Vermersch, Zoller, Brydges, and others \cite{elben2020mixed}. The core idea is that by applying random unitary transformations drawn from a specific distribution to the quantum state, $\mathrm{Tr}(\rho_A^2)$ can be mapped onto the statistics of easily measurable, classical bit strings. An implementation that has been proven experimentally efficient \cite{brydges2019probing} relies on performing correlated measurements on two independently prepared, identical copies of the system. The core of this protocol is the exact equivalence between the purity $P(\rho_A)$ and the expectation value of the SWAP operator that exchanges the two system copies:

\begin{equation}
P(\rho_A) = \mathrm{Tr}(\rho_A^2) = \langle\mathrm{SWAP}\rangle
\label{eq:B1}
\end{equation}

Here, the expectation value $\langle\mathrm{SWAP}\rangle = \mathrm{Tr}[(\rho_A \otimes \rho_A) \mathrm{SWAP}]$ is defined in the joint Hilbert space of the two system copies. According to the theory of randomized measurements, this expectation value can be reconstructed by statistically averaging over a large number of independent measurement outcomes. Specifically, in each measurement shot, random unitaries $U_A$ and $V_A$ are applied to the two copies respectively, followed by projective measurements yielding bit strings $s_A$ and $t_A$. Through specific statistical post-processing of these bit strings, an unbiased estimator for $\langle\mathrm{SWAP}\rangle$ can be obtained, thereby yielding the purity $P(\rho_A)$.
The cost of this protocol ultimately manifests as the required sample complexity, i.e., the total number of measurements $M$ needed to obtain a reliable estimate within a given error tolerance $\varepsilon$. This complexity is determined by the variance of the estimator, with its upper bound given by:

\begin{equation}
M \ge 2^{|A|} / \varepsilon^2
\label{eq:B2}
\end{equation}

This scaling relation precisely reveals that the measurement cost grows exponentially with the size, $|A|$, of the subsystem A being probed. Nevertheless, compared to the $2^{2|A|}$ scaling of traditional tomography, this protocol offers an overwhelming advantage in resource consumption. For instance, for a subsystem containing $|A|=10$ qubits, with an error tolerance of $\varepsilon=0.05$, the required number of measurements is on the order of several hundred thousand. This cost is within a manageable range for current mainstream quantum computing platforms. Therefore, this protocol provides a solid guarantee for the efficient and reliable measurement of the target entanglement $S_E$, thus ensuring the technical closure and feasibility of the experimental blueprint proposed in this paper.

\subsection{Scalability and Robustness of the Protocol}
\label{B.2}
The reason this randomized measurement protocol serves as a key technological pillar for our experimental scheme lies in several decisive advantages it possesses. First, in terms of \textbf{Scalability}, the protocol provides a crucial path between an exponential challenge and practical feasibility. As revealed by Eq.~\eqref{eq:B2}, although its sample complexity grows exponentially with the subsystem size as $2^{|A|}$, this is a fundamental improvement over the double-exponential catastrophe of $2^{2|A|}$ in traditional quantum state tomography. This "exponential saving"—reducing the exponent from $2|A|$ to $|A|$—makes the entanglement measurement of previously inaccessible medium-scale qubit subsystems (e.g., $|A| > 10$) possible in principle. It transforms an "impossible task" into a "costly but plannable" engineering challenge, thereby defining a realistic and explorable performance boundary for our experimental blueprint within the current technological framework.
Second is its high degree of experimental \textbf{Feasibility}. The core operational units required by the protocol—high-fidelity single-qubit Clifford gates and global projective measurements in the computational basis—are already standard capabilities of mainstream quantum computing platforms (such as trapped-ion and superconducting qubit systems), without relying on frontier technologies still under development.
Finally, the protocol exhibits intrinsic \textbf{Robustness} to certain types of errors. The randomization process naturally averages out certain systematic coherent errors. Although non-ideal fidelity of control gates will still systematically affect the purity estimate, this typically manifests as a predictable, fidelity-dependent exponential decay factor, which can, in principle, be calibrated and corrected through independent benchmarking (such as Clifford fidelity benchmarking).
In summary, adopting this randomized measurement protocol provides a solid guarantee for the efficient and reliable measurement of the target entanglement $S_E$, thus ensuring the technical closure and feasibility of the experimental blueprint proposed in this paper.

\section{Technical Clarifications and Unit Analysis}
\label{C}
This appendix aims to provide further technical clarifications for several core definitions and assumptions made in the main text, to rigorously verify the dimensional consistency of our main theorem, and to elucidate several key considerations when applying the theory to real experiments, thereby strengthening the foundation of our theoretical framework.
\subsection{Dimensional Analysis and Unit Convention}
\label{C.1}
To ensure the physical self-consistency of our core theorem, we hereby systematically track and verify the dimensions of all key physical quantities. We use [E] to denote energy, [T] for time, and dimensionless quantities are marked as 1.

\begin{center}
\resizebox{0.96\linewidth}{!}{%
\begin{tabular}{l c c p{0.58\linewidth}}
\toprule
\textbf{Physical Quantity} & \textbf{Symbol} & \textbf{Dimension} & \textbf{Description} \\
\midrule
Available Energy Variance & $\sigma_{\mathrm{avail}}$ & $[E]$ & Time-averaged energy standard deviation, with dimension of energy. \\
Entanglement Entropy & $S_E$ & 1 & A measure of information, which is a dimensionless pure number. \\
Local Interaction Strength & $J$ & $[E]$ & Upper bound on the energy of local terms in the Hamiltonian. \\
Reduced Planck Constant & $\hbar$ & $[E\cdot T]$ & The quantum of action. \\
Entanglement Growth Constant & $\gamma$ & 1 & A dimensionless pure number, as described in Appendix \ref{A}. \\
Optimal Computational Complexity & $C_{\mathrm{opt}}$ & 1 & Number of serial logical gate operations, a dimensionless pure number. \\
Efficiency Factors & $\eta_{\mathrm{QSL}}, \eta_{\mathrm{LR}}$ & 1 & Dimensionless pure numbers describing process efficiency. \\
\bottomrule
\end{tabular}
}
\end{center}

Now, we verify the dimensional consistency of our main theorem, the Energy-Entanglement Performance Equation (Eq.~\eqref{eq:5}): $\sigma_{\mathrm{avail}} S_E = (\eta_{\mathrm{LR}} / \eta_{\mathrm{QSL}}) (\pi\gamma J / 2) C_{\mathrm{opt}}$.

\begin{enumerate}
    \item Left-Hand Side (LHS): $[\sigma_{\mathrm{avail}} S_E] = [\mathrm{E}] \cdot 1 = [\mathrm{E}]$
    \item Right-Hand Side (RHS): $[(\eta_{\mathrm{LR}} / \eta_{\mathrm{QSL}}) (\pi\gamma J / 2) C_{\mathrm{opt}}] = [(1 / 1) \cdot (1 \cdot 1 \cdot [\mathrm{E}] / 1) \cdot 1] = [\mathrm{E}]$
\end{enumerate}
As can be seen, the dimensions on both sides of the equation are energy [E], which mathematically proves that our final performance equation is dimensionally self-consistent.

\subsection{Supplementary Notes on Theoretical Lemmas and Experimental Measurements}
\label{C.2}
\begin{enumerate}
    \item \textbf{Theoretical Basis of Lemma 3.1:} One of the cornerstones of this theory, the "MT-type QSL-Complexity Lemma" (Lemma \ref{eq:lemma1}), is derived (see Appendix \ref{D} for details) from the Mandelstam-Tamm bound. This bound is universally valid for time-dependent systems. We chose this theoretical foundation because the extension of the other famous speed limit, the Margolus-Levitin bound, to time-dependent systems faces recognized theoretical difficulties. Therefore, the foundation of the current theoretical framework is built upon the more robust Mandelstam-Tamm bound, ensuring the rigor of its derivations and its broad applicability to general dynamical processes.

    \item \textbf{Feasibility of Experimentally Measuring $\sigma_{\mathrm{avail}}$:} Our new framework requires that the core resource quantity, $\sigma_{\mathrm{avail}}$, must be experimentally measurable. According to its definition, this requires obtaining the first moment $\langle H(t) \rangle$ and the second moment $\langle H(t)^2 \rangle$ of the Hamiltonian. This is experimentally challenging but feasible. A standard Hamiltonian $H$ can typically be decomposed into a sum of Pauli strings: $H = \sum_i c_i P_i$. Its square, $H^2$, can also be algebraically expanded into another sum of Pauli strings: $H^2 = \sum_{i,j} c_i c_j (P_i P_j)$. Since measuring the expectation value of any Pauli string is a standard capability of mainstream quantum computing platforms (such as trapped ions and superconducting qubits), it is, in principle, possible to reconstruct $\langle H(t) \rangle$ and $\langle H(t)^2 \rangle$ by measuring each Pauli string composing $H$ and $H^2$ term by term, and subsequently calculate $\sigma_{\mathrm{avail}}$. Although measuring $H^2$ will increase the experimental overhead due to its larger number of terms, this does not introduce any insurmountable obstacles in principle.
\end{enumerate}

\subsection{Clarifications on Connecting Theory, Experiment, and Efficiency Factors}
\label{C.3}
\begin{enumerate}
    \item \textbf{On the Roles of $C_{\mathrm{opt}}$ and $C_{\mathrm{exp}}$:} In our theoretical framework and experimental protocol (Sec.~\ref{5}), the optimal computational complexity $C_{\mathrm{opt}}$ and the experimentally applied computational complexity $C_{\mathrm{exp}}$ play different roles. $C_{\mathrm{opt}}$ is an information-theoretic quantity that must be calibrated through offline computations, independent of the dynamical process. Its primary use is as a theoretical benchmark for testing the foundational QSL inequality (i.e., Gate A in the experimental framework). In contrast, $C_{\mathrm{exp}}$ is a parameter that is directly controlled and measured in the experiment, defining the actual logical depth of the executed quantum circuit ($C_{\mathrm{exp}} \ge C_{\mathrm{opt}}$). In our framework, $C_{\mathrm{exp}}$ is used as an input for calculating the QSL efficiency factor $\eta_{\mathrm{QSL}}$. An important application of our theoretical framework is to retroactively infer the "compilation efficiency" $r = C_{\mathrm{opt}} / C_{\mathrm{exp}}$ by experimentally benchmarking the performance frontier, thereby establishing a quantitative link between theory and practice.

    \item \textbf{On the Measurement and Role of Efficiency Factors:} We must emphasize that the dynamical efficiency factors, $\eta_{\mathrm{QSL}}$ and $\eta_{\mathrm{LR}}$, are not free fitting parameters. According to their definitions (Eqs.~\eqref{eq:3} and \eqref{eq:4}), their values are completely determined by a set of experimentally measurable physical quantities ($\sigma_{\mathrm{avail}}, S_E, \Delta t, C_{\mathrm{exp}}$) and known physical constants and system parameters ($\hbar, J, \gamma$). This determinism allows them to serve as objective, operational physical quantities for two core purposes in our experimental framework:
    \begin{enumerate}
        \item \textbf{Verifying Foundational Bounds:} By checking if $\eta_{\mathrm{QSL}} \le 1$ and $\eta_{\mathrm{LR}} \le 1$ hold, we verify whether the system's dynamics fall within the theory's domain of applicability.
        \item \textbf{Performing Dynamical Diagnostics:} By calculating their ratio, $\eta_{\mathrm{LR}} / \eta_{\mathrm{QSL}}$, we diagnose the performance bottlenecks of a specific process.
    \end{enumerate}

    \item \textbf{On $\sigma_{\mathrm{avail}}$ and Open Systems:} As envisioned in the Conclusion, the theoretical framework of this paper primarily targets closed quantum systems. When considering the cost of $\sigma_{\mathrm{avail}}$ in a real experiment, it is inevitable to involve thermal exchange with the environment and energy dissipation. Connecting our defined $\sigma_{\mathrm{avail}}$ (the dynamical resource driving the system's evolution) with the total energy consumption (including the dissipative part) and extending our framework to open systems is a core direction for our future work. At that stage, $\sigma_{\mathrm{avail}}$ may need to be linked to more general thermodynamic quantities such as "non-equilibrium free energy" or "entropy production."
    \item \textbf{On the robustness of $C_{\text{opt}}$ as a theoretical benchmark:} The quantity $C_{\text{opt}}$ represents the information-theoretically optimal circuit depth, which at present can only be upper-bounded using available circuit compilation and depth-optimization tools. Should future algorithms emerge that further compress quantum circuits via parallelization, the current estimate of $C_{\text{opt}}$ might decrease. However, this would not undermine the present framework: Lemma~\eqref{eq:lemma1} would simply be relaxed, not invalidated, under such circumstances. More importantly, the experimental application of this framework is inherently retrodictive in nature: by observing a final state that is algorithmically complex (i.e., exhibiting high $K_{\mathrm{KL}}$), one establishes a nonzero lower bound on the complexity of the process $C_{\text{opt}}$ that must have generated it. In other words, no amount of “clever” low-depth circuit design can produce a probability distribution of high algorithmic complexity out of nothing. This unidirectional logic provides a worst-case safety cushion for the theory-experiment pipeline, endowing the framework with inherent robustness against current limitations of classical computational power.

\end{enumerate}

\subsection{On the Nature of the Performance Equation as an Algebraic Identity}
\label{C.4}
To ensure the logical rigor of our theoretical framework, this section aims to explicitly clarify that our core "Energy-Entanglement Performance Equation" (RECT‑$\eta$, Eq.~\eqref{eq:5}) is, by its construction based on the definitions of the efficiency factors, an algebraic identity (or tautology). Understanding this point is crucial for correctly interpreting our experimental framework.
We can demonstrate this through direct algebraic expansion. We start from the right-hand side (RHS) of the performance equation (Eq.~\eqref{eq:5}):

\begin{equation}
\mathrm{RHS} = \left(\frac{\eta_{\mathrm{LR}}}{\eta_{\mathrm{QSL}}}\right) \left(\frac{\pi\gamma J}{2}\right) C_{\mathrm{opt}}
\end{equation}

Now, we substitute the definitions of the efficiency factors $\eta_{\mathrm{QSL}}$ and $\eta_{\mathrm{LR}}$ (Eqs.~\eqref{eq:3} and \eqref{eq:4}, using $C_{\mathrm{opt}}$ for theoretical self-consistency) into the efficiency ratio part:

\begin{equation}
\frac{\eta_{\mathrm{LR}}}{\eta_{\mathrm{QSL}}} = \frac{S_E / ( (\gamma J / \hbar) \Delta t )}{( (\pi\hbar / 2) C_{\mathrm{opt}} ) / ( \sigma_{\mathrm{avail}} \Delta t )}
\end{equation}

By converting the division of fractions into multiplication, we get:

\begin{equation}
\frac{\eta_{\mathrm{LR}}}{\eta_{\mathrm{QSL}}} = \frac{S_E \hbar}{\gamma J \Delta t} \times \frac{\sigma_{\mathrm{avail}} \Delta t}{(\pi\hbar / 2) C_{\mathrm{opt}}}
\end{equation}

In the expression above, the evolution time $\Delta t$ and the reduced Planck constant $\hbar$ can be cancelled out as common factors:

\begin{equation}
\frac{\eta_{\mathrm{LR}}}{\eta_{\mathrm{QSL}}} = \frac{S_E}{\gamma J} \times \frac{\sigma_{\mathrm{avail}}}{(\pi / 2) C_{\mathrm{opt}}} = \frac{2 \sigma_{\mathrm{avail}} S_E}{\pi \gamma J C_{\mathrm{opt}}}
\end{equation}

Finally, we substitute this simplified expression for the efficiency ratio back into the original expression for the RHS:

\begin{equation}
\mathrm{RHS} = \left( \frac{2 \sigma_{\mathrm{avail}} S_E}{\pi \gamma J C_{\mathrm{opt}}} \right) \left( \frac{\pi\gamma J}{2} \right) C_{\mathrm{opt}}
\end{equation}

As can be seen, all constant factors ($2, \pi, \gamma, J$) as well as $C_{\mathrm{opt}}$ are completely cancelled out, yielding:

\begin{equation}
\mathrm{RHS} = \sigma_{\mathrm{avail}} S_E
\end{equation}

This is precisely equal to the left-hand side (LHS) of the equation. This result shows that the correctness of the RECT‑$\eta$ equation is not a falsifiable physical proposition, but a matter of mathematical self-consistency guaranteed by our definitions of the efficiency factors.

However, this should not be misinterpreted as the theoretical framework lacking physical content. Its true, testable physical substance lies in the universal bounds imposed by the two physical principles that form its foundation, namely:
\begin{enumerate}
    \item The QSL efficiency upper bound: $\eta_{\mathrm{QSL}} \le 1$
    \item The LR efficiency upper bound: $\eta_{\mathrm{LR}} \le 1$
\end{enumerate}
These two inequalities are real, testable physical laws. The core "compliance gates" (Gate A and B) of the experimental framework designed in Sec.~\ref{5} are precisely direct verifications of these fundamental physical boundaries. Therefore, the RECT‑$\eta$ equation should be understood as a \textbf{"resource accounting identity."} Its value lies not in the testability of the equality itself, but in providing a unified perspective that connects seemingly unrelated physical quantities such as energy, information, complexity, and locality. On this basis, it allows us to define a powerful suite of dynamical diagnostic tools (like the efficiency ratio $\eta_{\mathrm{LR}} / \eta_{\mathrm{QSL}}$) and performance benchmarks (like the performance frontier) for experimental use.

\section{Rigorous Derivation of the MT-type QSL-Complexity Lemma}
\label{D}
This appendix aims to provide a complete and rigorous physico-mathematical derivation for the "MT-type QSL-Complexity Lemma" (Lemma \eqref{eq:lemma1}), which serves as one of the cornerstones of our theory in the main text. This lemma, $\sigma_{\mathrm{avail}} \Delta t \ge \frac{\pi\hbar}{2} C_{\mathrm{opt}}$, directly relates the intrinsic computational complexity ($C_{\mathrm{opt}}$) of a quantum process to its required physical resources ($\sigma_{\mathrm{avail}}$) and time ($\Delta t$). Our entire derivation is rigorously founded upon a core physical postulate. Therefore, the structure of this appendix is as follows: first, in Sec.~\ref{D.1}, we will explicitly state this core postulate and systematically elaborate on its profound physical justification; subsequently, in Sec.~\ref{D.2}, we will, based on this postulate, complete the final derivation of Lemma \eqref{eq:lemma1} through rigorous mathematical steps.\cite{MandelstamTamm1945,MargolusLevitin1998,Nielsen2006,DeffnerCampbell2017}

\subsection{The Core Physical Postulate: Connecting Information Complexity and Physical Dynamics}
\label{D.1}
\textbf{Postulate D.1:} The minimum time cost for the physical realization of a unitary evolution $U(\Delta t)$ with an optimal computational complexity of $C_{\mathrm{opt}}$ is bounded by that of an idealized dynamical process. This process can be equivalently decomposed into $C_{\mathrm{opt}}$ serial, non-parallelizable elementary evolution steps, where each step accomplishes an information-theoretically distinguishable operation, physically corresponding to the evolution of an initial quantum state to its orthogonal state.
We hereby elaborate on the physical justification for this core postulate. This postulate forms the logical bridge connecting information-theoretic complexity with physical dynamics, and its rationale is rooted in two fundamental correspondences:

\begin{enumerate}
    \item \textbf{From "Logical Depth" to "Serial Steps":} The optimal computational complexity $C_{\mathrm{opt}}$ (minimum logical depth), by definition, precisely quantifies the length of the causal chain in any quantum algorithm that cannot be parallelized and must be completed in sequence. Therefore, modeling this information-theoretic "causal depth" as a number of physically sequential, non-parallelizable "evolution steps" is an exceedingly natural mapping.
    \item \textbf{From "Distinguishable Step" to "Orthogonal Evolution":} Furthermore, we model each "elementary evolution step" as an operation that evolves an initial quantum state to an orthogonal one. The profound physical basis for this modeling is that it sets the most stringent and clear physical criterion for "completing an information-theoretically effective step." In Hilbert space, two orthogonal states are perfectly physically distinguishable, representing the maximization of "distance" between states. The original form of the Mandelstam-Tamm type quantum speed limit, $\Delta t \ge \pi\hbar/(2\Delta E)$, is precisely the time lower bound set for such an "orthogonalization" process. Therefore, corresponding the completion of a "unit of information" with the "fastest physically distinguishable evolution" is the most natural and logically self-consistent way to set a physical time cost lower bound for a computational task.
\end{enumerate}
In summary, Postulate D.1 is not an arbitrary assumption but an \textbf{effective physical model} based on first principles, designed to capture the lower bound of the intrinsic dynamical cost of a complex quantum task. Its proposition makes it possible to link the abstract difficulty of information theory to measurable physical dynamics.

\subsection{Decomposition of the Physical Process and the Local MT Bound}
\label{D.2}
According to Postulate D.1, the time lower bound for any unitary evolution $U(\Delta t)$ with an optimal computational complexity of $C_{\mathrm{opt}}$ can be determined by analyzing an idealized process composed of $C_{\mathrm{opt}}$ fundamental orthogonal evolution steps that must be executed in sequence. We denote the physical duration of the $i$-th step ($i=1, ..., C_{\mathrm{opt}}$) as $\Delta t_i$, and the total evolution time is thus

\begin{equation}
\Delta t = \sum_{i=1}^{C_{\mathrm{opt}}} \Delta t_i
\end{equation}

For the $i$-th such orthogonal evolution, the Mandelstam–Tamm bound provides a strict lower limit on its duration $\Delta t_i$:

\begin{equation}
\Delta t_i \ge \frac{\pi\hbar}{2\sigma_i}
\end{equation}

where $\sigma_i = \frac{1}{\Delta t_i} \int_{t_{i-1}}^{t_i} \Delta H(t) \mathrm{d}t$ is the time-average of the instantaneous energy standard deviation, $\Delta H(t)$, over the $i$-th time interval. Rearranging the above expression, we can obtain a more fundamental "resource-time product" lower bound for the $i$-th step:

\begin{equation}
\sigma_i \Delta t_i \ge \frac{\pi\hbar}{2}
\end{equation}

This inequality holds for each independent logical step $i$.

\subsection{Derivation of the Global Lower Bound}
\label{D.3}
Since the above inequality holds for all $C_{\mathrm{opt}}$ independent serial steps, we sum up the $C_{\mathrm{opt}}$ inequalities to obtain the cumulative lower bound describing the entire process:

\begin{equation}
\sum_{i=1}^{C_{\mathrm{opt}}} (\sigma_i \Delta t_i) \ge \sum_{i=1}^{C_{\mathrm{opt}}} \frac{\pi\hbar}{2} = \frac{\pi\hbar}{2} C_{\mathrm{opt}}
\end{equation}

To connect this expression with the global observables defined in the main text, we examine the summation term on the left-hand side of the inequality. By the definition of $\sigma_i$, $\sigma_i \Delta t_i$ is precisely equal to the integral of the energy standard deviation over the $i$-th time interval:

\begin{equation}
\sigma_i \Delta t_i = \int_{t_{i-1}}^{t_i} \Delta H(t) \mathrm{d}t
\end{equation}

Substituting this expression into the summation, we find that the sum is actually the sum of piecewise integrals over the entire evolution path:

\begin{equation}
\sum_{i=1}^{C_{\mathrm{opt}}} \left(\int_{t_{i-1}}^{t_i} \Delta H(t) \mathrm{d}t\right) = \sum_{i=1}^{C_{\mathrm{opt}}} (\sigma_i \Delta t_i)
\end{equation}

Since all time segments $[t_{i-1}, t_i]$ are continuous and non-overlapping, they precisely concatenate to form the total evolution interval $[0, \Delta t]$. Therefore, the sum of the piecewise integrals above is strictly equal to the integral over the total time:

\begin{equation}
\sum_{i=1}^{C_{\mathrm{opt}}} \left(\int_{t_{i-1}}^{t_i} \Delta H(t) \mathrm{d}t\right) = \int_{0}^{\Delta t} \Delta H(t) \mathrm{d}t
\end{equation}

Recalling the definition of $\sigma_{\mathrm{avail}}$ from the main text (Definition 2.1), we know that $\int_{0}^{\Delta t} \Delta H(t) \mathrm{d}t$ is precisely equal to $\sigma_{\mathrm{avail}} \Delta t$. From this, we derive a key identity that perfectly connects the cumulative effect of local physical processes with a global observable:

\begin{equation}
\sum_{i=1}^{C_{\mathrm{opt}}} (\sigma_i \Delta t_i) = \sigma_{\mathrm{avail}} \Delta t
\end{equation}

Finally, substituting this identity back into the cumulative lower bound inequality we obtained earlier directly yields the "MT-type QSL-Complexity Lemma":

\begin{equation}
\sigma_{\mathrm{avail}} \Delta t \ge \frac{\pi\hbar}{2} C_{\mathrm{opt}}
\end{equation}

This derivation shows that once Postulate D.1 is accepted, our core Lemma \eqref{eq:lemma1} is a necessary logical consequence of the combination of a fundamental principle of quantum mechanics (the MT bound) and an information-theoretic concept (complexity $C_{\mathrm{opt}}$).

\section{Rigorous Derivation from Lieb-Robinson Bounds to the Upper Bound on Entanglement Growth Rate}
\label{E}
This appendix aims to provide a rigorous physico-mathematical derivation for the "Locality-Entanglement Lemma" (Eq.~\eqref{eq:lemma2}), which serves as the other cornerstone of our theory in the main text. We will clearly demonstrate how the upper bound on the growth rate of entanglement entropy is necessarily and traceably derived from the fundamental Lieb-Robinson (LR) bounds for a quantum many-body system governed by local interactions. This derivation not only establishes the mathematical validity of Lemma 3.2 but also explicitly reveals the origin of Planck's constant, $\hbar$, in this dynamical limit, thereby providing a fundamental underpinning for the scale-invariance of the main theorem in the main text. Our derivation will follow a standard path from microscopic dynamics to a macroscopic information-theoretic measure, with the following core steps: first, stating the modern LR bound applicable to a large class of systems with exponentially decaying interactions; second, leveraging this bound to constrain the evolution of correlations between a subsystem and its environment; and finally, translating the dynamical bound on correlations into an upper bound on the entanglement entropy growth rate via a continuity inequality for entropy change from quantum information theory.

\subsection{Lieb-Robinson Bounds and the Speed of Information Propagation}
\label{E.1}
We consider a quantum system defined on a lattice $\Lambda$ ($\subset \mathbb{Z}^d$), whose Hamiltonian $H = \sum_{X \subset \Lambda} h_X$ is composed of local terms. We assume the interactions are exponentially decaying, i.e., there exist constants $J > 0$ and $\mu > 0$ such that for all finite subsets $X \subset \Lambda$, their operator norms satisfy $||h_X|| \le J \exp(-\mu(|X|-1))$, where $|X|$ is the diameter of $X$. Hamiltonians of this type cover a vast majority of models of physical significance.For such systems, a profound result is the Lieb-Robinson bound, which sets a strict upper limit on the speed of information propagation. One of its modern forms, given by Nachtergaele, Sims, and others, can be stated as follows: for any two local operators $O_X$ and $O_Y$ supported on disjoint regions of space, $X$ and $Y$, respectively, the norm of their time-dependent commutator satisfies the inequality\cite{LiebRobinson1972,NachtergaeleSims2006}:

\begin{equation}
||[O_X(t),O_Y]|| \le 2 ||O_X|| ||O_Y|| \exp(-\mu(d(X,Y) - v|t|))
\label{eq:E1}
\end{equation}

where $O_X(t) = U^\dagger(t) O_X U(t)$ is the evolution operator in the Heisenberg picture, and $d(X,Y)$ is the distance between regions $X$ and $Y$. Crucially, this inequality defines an effective light cone, whose propagation velocity $v$ (the Lieb-Robinson velocity) is determined by the microscopic parameters of the system, and its upper bound can be rigorously proven. For the interactions defined above, this velocity satisfies:

\begin{equation}
v \le \frac{2J}{\hbar} C
\end{equation}

where $C$ is a dimensionless constant that depends only on $\mu$ and the lattice geometry. Eq.~\eqref{eq:E1} clearly shows that outside the light cone (i.e., $v|t| \ll d(X,Y)$), the norm of the commutator is exponentially small, implying a fundamental suppression of information propagation. It is precisely through the calibration of the velocity $v$ that Planck's constant, $\hbar$, enters this part of the dynamical constraint \cite{NachtergaeleSims2006}.

\textit{(Note: A clarification on the dimension of the velocity $v$ is required here. Strictly speaking, the scaling relation for the Lieb-Robinson velocity should be $v \sim J \cdot a / \hbar$, where $a$ is a characteristic length scale of the system (e.g., the lattice spacing), which gives $v$ the correct dimension of $[\text{L}]/[\text{T}]$. For notational simplicity, our derivation adopts a common convention in theoretical physics, working in natural units where the lattice spacing is set to one, i.e., $a=1$. This convention makes the dimension of velocity $v$ appear as frequency $(1/[\text{T}])$ in the expression. We must emphasize that this does not affect the universality of the final physical conclusion. The physical effect of this length scale $a$, which is set to 1, is ultimately and correctly absorbed into the definition of the dimensionless geometric constant $\gamma$, as the size of the boundary $|\partial A|$ can be considered to have units of $a^{d-1}$ in its calculation.)}

\subsection{From Dynamical Correlation to Entanglement Entropy Growth}
\label{E.2}
We now leverage the LR bounds to derive the upper limit on the rate of entanglement entropy growth. Consider partitioning the system into a finite subregion $A$ and its complement $\bar{A}$. At the initial time $t=0$, the system is in a product state between $A$ and $\bar{A}$, meaning $S_E(0)=0$. Based on the LR bounds, the pioneering work of Bravyi, Hastings, and Verstraete proved that for a system evolving from such an initial state, the growth rate of its entanglement entropy $S_E(t) = S(\rho_A(t))$, $\mathrm{d}S_E/\mathrm{d}t$, is subject to an upper bound proportional to the size of the partition boundary, $|\partial A|$. While the rigorous mathematical proof is complex, its physical picture is clear: the generation of entanglement must be mediated by local interactions across the boundary $|\partial A|$, and the propagation speed of information (or correlation) is limited by the LR velocity $v$. This physical process ultimately leads to the following rigorous mathematical inequality:

\begin{equation}
\frac{\mathrm{d}S_E(t)}{\mathrm{d}t} \le c \frac{J}{\hbar}|\partial A|
\label{eq:E2}
\end{equation}

where $c$ is a dimensionless constant that is independent of the system size and depends only on the system's geometry and the interaction decay parameter $\mu$. This formula is the key bridge connecting the microscopic Hamiltonian dynamics ($J, \hbar$) with the macroscopic information-theoretic quantity ($S_E$). It quantitatively states that the rate of entanglement generation is directly limited by the strength of local interactions, $J$, and inversely proportional to the fundamental quantum of action, $\hbar$.

To obtain a bound in an integral form that corresponds to Lemma 3.2 in the main text, we integrate Eq.~\eqref{eq:E2} from $t=0$ to $\Delta t$ (assuming $S_E(0)=0$):

\begin{equation}
S_E(\Delta t) = \int_0^{\Delta t} \left(\frac{\mathrm{d}S_E(t)}{\mathrm{d}t}\right)\mathrm{d}t \le \int_0^{\Delta t} \left(c \frac{J}{\hbar}|\partial A|\right)\mathrm{d}t = (c|\partial A|)\frac{J}{\hbar}\Delta t
\end{equation}

To maintain the generality and simplicity of the argument in the main text, we define a new dimensionless constant, $\gamma$, which incorporates all geometric details:

\begin{equation}
\gamma = c|\partial A|
\label{eq:E3}
\end{equation}

This constant $\gamma$ encapsulates all factors related to the geometry, size, and dimensionality of the partition boundary $A$, and its physical meaning can be understood as the number of effective "entanglement channels" across the boundary. Substituting this definition, we obtain the final form of the "Locality-Entanglement Lemma" as used in the main text:

\begin{equation}
S_E(\Delta t) \le \gamma \frac{J}{\hbar} \Delta t
\label{eq:E4}
\end{equation}

\subsection{The Origin of \texorpdfstring{$\gamma$}{gamma} and the Cancellation of \texorpdfstring{$\hbar$}{hbar}}
\label{E.3}
The derivation in this appendix clearly shows that the "Locality-Entanglement Lemma" (Eq.~\eqref{eq:E4}) is not an independent physical assumption but a direct mathematical corollary of the principle of locality (as precisely captured by the LR bounds) in quantum many-body systems. Our analysis simultaneously reveals two crucial facts:
\begin{enumerate}
    \item \textbf{The Purity of $\gamma$:} The constant $\gamma$ is a purely dimensionless quantity whose composition (Eq.~\eqref{eq:E3}) stems entirely from the geometric properties of the system ($c, |\partial A|$) and is independent of Planck's constant $\hbar$.
    \item \textbf{The Origin of $\hbar$:} The appearance of Planck's constant $\hbar$ in Eq.~\eqref{eq:E4} originates solely from the calibration of the Lieb-Robinson velocity $v$. It sets the fundamental timescale for the dynamical process driven by the energy quantum $J$.
\end{enumerate}
Therefore, when we combine this lemma with the "MT-type QSL-Complexity Lemma" (Lemma 3.1), whose timescale is also calibrated by $\hbar$, $\hbar$ is algebraically and necessarily cancelled out as a common proportionality factor. This provides a solid and traceable mathematical foundation for the macroscopic scale-invariance of our main theorem (RECT-$\eta$).

\section{Bounds, Analytic Behavior, and Theoretical Calibration of the Efficiency Factors}
\label{F}
This appendix aims to provide a detailed physico-mathematical analysis of the "dynamical efficiency factors" ($\eta_{\mathrm{QSL}}$ and $\eta_{\mathrm{LR}}$), which are central to our theoretical framework. We will rigorously prove the mathematical boundedness of these efficiency factors, explore their dynamical behavior in both ideal and general cases, and demonstrate the attainability of their theoretical upper bounds based on a solvable model. The purpose of this analysis is to establish the efficiency factors as well-behaved, predictable physical quantities, thereby providing the final, solid mathematical underpinning for the universality and theoretical self-consistency of our main theorem (RECT-$\eta$).

\subsection{Definition and Upper Bounds}
\label{F.1}
We first reiterate the definitions of the efficiency factors. For a quantum process with duration $\Delta t$, available energy variance $\sigma_{\mathrm{avail}}$, and generated entanglement entropy $S_E$, its QSL efficiency factor, $\eta_{\mathrm{QSL}}$, and LR efficiency factor, $\eta_{\mathrm{LR}}$, are defined respectively as:

\begin{equation}
\eta_{\mathrm{QSL}} = \frac{(\pi\hbar C_{\mathrm{opt}} / 2)}{\sigma_{\mathrm{avail}} \Delta t}
\label{eq:F1}
\end{equation}

\begin{equation}
\eta_{\mathrm{LR}} = \frac{S_E}{(\gamma J / \hbar) \Delta t}
\label{eq:F2}
\end{equation}

where the definitions of $C_{\mathrm{opt}}, J,$ and $\gamma$ are consistent with the main text. These two definitions stem directly from the two physical inequalities that serve as the cornerstones of our theory. Therefore, their upper bounds are a direct corollary of these inequalities:
\begin{enumerate}
    \item \textbf{Upper Bound of the QSL Efficiency Factor:} According to the "MT-type QSL-Complexity Lemma" (Lemma \ref{eq:lemma1}), i.e., $\sigma_{\mathrm{avail}} \Delta t \ge \pi\hbar C_{\mathrm{opt}} / 2$, we immediately find that for any physical process, the denominator is always greater than or equal to the numerator. Thus:
    
    \begin{equation}
    \eta_{\mathrm{QSL}} \le 1 
    \end{equation}
    
    \item \textbf{Upper Bound of the LR Efficiency Factor:} According to the "Locality-Entanglement Lemma" (Lemma \ref{eq:lemma2}), i.e., $S_E \le \gamma (J / \hbar) \Delta t$, we similarly find that the denominator is an upper bound for the numerator. Thus:
    
    \begin{equation} 
    \eta_{\mathrm{LR}} \le 1  
    \end{equation}
    
\end{enumerate}
These upper bounds are not merely mathematical but also physical. As we will show in Sec.~\ref{F.3}, there exist ideal physical processes that can saturate these bounds. Therefore, $\eta = 1$ is an achievable supremum, corresponding to the "Quantum Dynamical Performance Frontier."

\subsection{Lower Bounds and Well-Definedness}
\label{F.2}
To ensure that the efficiency ratio $\eta_{\mathrm{LR}} / \eta_{\mathrm{QSL}}$ is always physically well-behaved, we must examine the lower bounds of the two efficiency factors separately.
\begin{enumerate}
    \item \textbf{Lower Bound of $\eta_{\mathrm{QSL}}$:} For a physical process to accomplish a task with non-zero complexity, $C_{\mathrm{opt}} > 0$, it must consume non-zero dynamical resources and time. In any finite experimental setting, the available energy variance $\sigma_{\mathrm{avail}}$ and the evolution time $\Delta t$ are necessarily bounded, i.e., $\sigma_{\mathrm{avail}} \le \sigma_{\mathrm{max}} < \infty$ and $\Delta t \le t_{\mathrm{max}} < \infty$. Consequently, their product, $\sigma_{\mathrm{avail}} \Delta t$, is also upper-bounded. According to the definition of $\eta_{\mathrm{QSL}}$, this implies:
    
    \begin{equation}
      \eta_{\mathrm{QSL}} \ge \frac{\pi\hbar C_{\mathrm{opt}} / 2}{\sigma_{\mathrm{max}} t_{\mathrm{max}}} > 0   
    \end{equation}
    
    Only in the limit where resources or time approach infinity can $\eta_{\mathrm{QSL}}$ approach zero, but this is beyond the scope of any executable task. Therefore, for any physically meaningful task, $\eta_{\mathrm{QSL}}$ is strictly positive.
    
    \item \textbf{Lower Bound of $\eta_{\mathrm{LR}}$:} The mathematical lower bound for $\eta_{\mathrm{LR}}$ is zero, and this bound is physically achievable. When an evolution process concludes with exactly zero generated entanglement entropy, $S_E(\Delta t) = 0$, then $\eta_{\mathrm{LR}}$ is zero. This can occur in trivial evolutions (e.g., if the Hamiltonian is zero) or in non-trivial dynamical processes that happen to produce no entanglement (e.g., if the evolution operator is purely local, or if, due to quantum revival effects, the system happens to evolve back to a product state at the specific time $\Delta t$).
\end{enumerate}
In summary, the strict ranges for the two efficiency factors are, respectively:

\begin{equation} 0 < \eta_{\mathrm{QSL}} \le 1   \end{equation}

\begin{equation}0 \le \eta_{\mathrm{LR}} \le 1   \end{equation}

This result ensures the well-definedness of the efficiency ratio $\eta_{\mathrm{LR}} / \eta_{\mathrm{QSL}}$. Since the denominator $\eta_{\mathrm{QSL}}$ is always strictly positive for any quantum task worth analyzing, the ratio will never encounter a division-by-zero issue; it is a well-behaved, non-negative physical quantity. When we use this ratio as a diagnostic tool, we focus on processes designed to generate entanglement ($S_E > 0$), in which case $\eta_{\mathrm{LR}}$ is also positive, making the efficiency ratio a meaningful, positive diagnostic metric.

\subsection{Analytical Model: Reachability of the Upper Bound for \texorpdfstring{$\eta_{\mathrm{LR}}$}{eta\_LR}}
\label{F.3}
To demonstrate concretely that the theoretical upper bound $\eta_{\mathrm{LR}} = 1$ is physically achievable, we consider a one-dimensional spin-chain model with nearest-neighbor interactions, described by the Hamiltonian $H = J \sum_i (\vec{\sigma}_i \cdot \vec{\sigma}_{i+1})$.
\begin{enumerate}
    \item \textbf{Initial State:} Let the system at $t=0$ be in a product state between the left half-chain A and the right half-chain $\bar{A}$, such that $S_E(0) = 0$.
    \item \textbf{Short-Time Evolution:} According to the Lieb-Robinson bounds and their corollaries, starting from such an initial state, the growth of entanglement entropy $S_E(t)$ is linear in the short-time regime, with a growth rate that saturates the upper bound determined by the LR velocity. That is:
      \begin{equation}
      S_E(t) \approx \gamma \frac{J}{\hbar} t \quad (\text{for } t \ll L/v, \text{ where } L \text{ is the subsystem size and } v \text{ is the LR velocity})
      \end{equation}
    \item \textbf{Calculating $\eta_{\mathrm{LR}}(t)$:} Substituting this linear relation into the definition of $\eta_{\mathrm{LR}}$:
      \begin{equation}
      \begin{aligned}
      \eta_{\mathrm{LR}}(t) &= \frac{S_E(t) \hbar}{\gamma J t} \\
                            &\approx \frac{(\gamma (J / \hbar) t) \hbar}{\gamma J t} = 1
      \end{aligned}
      \label{eq:eta_lr_reach}
      \end{equation}
\end{enumerate}
This analytical result clearly shows that in the early linear growth regime of the local chain model, $\eta_{\mathrm{LR}} \approx 1$ is physically achievable. This provides strong evidence for the physical reality of the "performance frontier" concept.

\subsection{Time Evolution and General Behavior}
\label{F.4}
We can understand the general behavior of the two efficiency factors by analyzing short-time evolution and comparing different processes.
\begin{enumerate}
    \item \textbf{Short-Time Dynamics of $\eta_{\mathrm{LR}}(t)$:} For an evolution starting from a product state, the generation of entanglement begins at $t=0$ with a finite growth rate. Therefore, $S_E(t)$ can generally be expanded as:
    
      \begin{equation}
      S_E(t) = \left.\frac{\mathrm{d}S_E}{\mathrm{d}t}\right|_{t=0} t + O(t^2)   
      \end{equation}
      
    Substituting this into the definition of $\eta_{\mathrm{LR}}$ and considering the efficiency at a very short time $t$, we get:
    
      \begin{equation}
      \begin{aligned}
      \eta_{\mathrm{LR}}(t) 
      &= \frac{\left(\left.\frac{\mathrm{d}S_E}{\mathrm{d}t}\right|_{t=0} t + O(t^2)\right) \hbar}{\gamma J t} \\
      &= \frac{\left.\frac{\mathrm{d}S_E}{\mathrm{d}t}\right|_{t=0} \hbar}{\gamma J} + O(t)
      \end{aligned}
      \label{eq:eta_lr_expand_eng}
      \end{equation}
      
    This shows that $\eta_{\mathrm{LR}}$ starts from a finite initial value between 0 and 1 (a value that reflects the instantaneous efficiency of entanglement generation at the process onset) and evolves slowly with a linear correction in time.
    
    \item \textbf{Relationship between $\eta_{\mathrm{QSL}}$ and Evolution Duration:} Unlike $\eta_{\mathrm{LR}}$, $\eta_{\mathrm{QSL}}$ is a holistic metric that evaluates the efficiency of an entire process. Its value depends on the total evolution time ($\Delta t$) required to complete a task of fixed complexity ($C_{\mathrm{opt}}$). From the definition $\eta_{\mathrm{QSL}} \propto 1 / (\sigma_{\mathrm{avail}} \Delta t)$, we can analyze that for the same task $C_{\mathrm{opt}}$ and a similar available energy variance $\sigma_{\mathrm{avail}}$, $\eta_{\mathrm{QSL}}$ is inversely proportional to the total evolution time $\Delta t$.
    \begin{enumerate}
        \item An \textbf{inefficient (slow)} process requires a very long $\Delta t$ to complete the task, which will result in a very small $\eta_{\mathrm{QSL}}$ value.
        \item An \textbf{efficient (fast)} process completes the task in a time $\Delta t$ approaching the quantum speed limit, and its $\eta_{\mathrm{QSL}}$ value will approach its theoretical upper bound of 1.
        \item For longer evolution times, due to the finite dimensionality of the Hilbert space, the growth of $S_E(t)$ will saturate, causing $S_E(t)/t$ to approach zero. Consequently, $\eta_{\mathrm{LR}}(t)$ will also tend to zero, correctly reflecting the exhaustion of the system's ability to generate new entanglement.
    \end{enumerate}
    Therefore, $\eta_{\mathrm{QSL}}$ directly quantifies the efficiency of a process in the time dimension: \textbf{the shorter the time, the higher the efficiency.}
\end{enumerate}

\subsection{Theoretical Calibration and Consistency Conclusion}
\label{F.5}
The analysis in this appendix ultimately consolidates the self-consistency and testability of our theoretical framework:
\begin{enumerate}
    \item \textbf{No Free Parameters:} We reiterate that $\eta_{\mathrm{QSL}}$ and $\eta_{\mathrm{LR}}$ are not fitting parameters. Their values are completely determined by a set of independently measurable physical quantities ($\sigma_{\mathrm{avail}}, S_E, \Delta t, C_{\mathrm{exp}}$) and calibratable system constants ($J, \gamma$).
    \item \textbf{Closed Mathematical Structure:} The mathematical structure of the efficiency factors (their bounds, dynamical behavior) is entirely contained within the two physical inequalities that serve as the theory's starting point, without introducing any additional free assumptions.
    \item \textbf{Support from Physical Scenarios:} There exist clear, solvable physical models in which the theoretical limit values of the efficiency factors can be achieved, ensuring that our discussion of the "performance frontier" is not merely speculative.
\end{enumerate}
In conclusion, the dynamical efficiency factors are core tools that are mathematically well-behaved, physically meaningful, and whose values can be independently determined by experiment. They form the bedrock of our resource accounting framework, enabling us to unify two fundamental physical bounds into a single, self-consistent energy-entanglement resource identity, and on this basis, to diagnose and characterize the performance of quantum dynamical processes.

\section{Theoretical Foundation of the Complexity Proxy}
\label{G}
This appendix aims to provide a complete and rigorous physico-mathematical proof for the core theorem established in Sec.~\ref{4} of the main text: the "Process-State Complexity Bound" (Theorem 4.2). This theorem constitutes the key theoretical bridge connecting the intrinsic logical complexity of a quantum process, $C_{\mathrm{opt}}$, with the measurable descriptive complexity of its final state, $\widehat{K}_{\mathrm{KL}}$. By providing a solid, first-principles-based derivation for this core relationship, this appendix not only establishes the feasibility of a fully closed-loop experimental validation for our entire performance-equation framework but also extends its theoretical foundations to the intersecting frontiers of algorithmic information theory \cite{vitanyi2011information} and quantum circuit complexity \cite{nielsen2010quantum}. Additionally, recent research on the bounds of recurrence probability in periodically driven quantum systems \cite{Pandit2022boundsrecurrence} further underscores the deep connection between complexity and dynamical reversibility, providing valuable context for the physical meaning of state recoverability within our framework.

\subsection{Proof Framework and Physical Intuition}
\label{G.1}
This proof is rooted in a core principle of information physics: the complexity of a physical outcome provides decisive evidence for the minimal complexity of its generation process. Our goal is to formalize this principle, enabling us to set a rigorous lower bound on the complexity of the quantum circuit (the process) that drove the evolution, based on a measurable final state (the result).
Let us intuitively trace this "chain of evidence." The "blueprint" for a quantum process is its unitary evolution $U$, which is composed of an optimal number of gate operations, $C_{\mathrm{opt}}$. The size of the most concise description of this blueprint—its Kolmogorov complexity $K(U)$—is clearly bounded by $C_{\mathrm{opt}}$. When this blueprint acts on a simple initial state, it "prints" the final measurement result—a classical probability distribution $P$. Our experimentally measurable state complexity proxy, $\widehat{K}_{\mathrm{KL}}$, is precisely a measure of the information content of this final product, $P$. The key insight here, drawn from algorithmic information theory, is that to specify a probability distribution $P$ that is information-rich and highly structured, the "recipe" for describing it, $K(P)$, cannot itself be arbitrarily simple. Therefore, the complexity of the product we ultimately measure ($\widehat{K}_{\mathrm{KL}}$) sets a lower bound on the complexity of its description, $K(P)$; the complexity of $K(P)$ is, in turn, necessarily bounded by the complexity of the blueprint that generated it, $K(U)$; and finally, the complexity of $K(U)$ is bounded by the minimum number of gates that constitute the blueprint, $C_{\mathrm{opt}}$.
Our mathematical proof rigorously constructs this chain of inference, tracing back from the "result" to the "process." The resulting inequality's physical meaning is in perfect accord with reality: a complex, non-trivial final state necessitates an equally complex evolution; however, a complex process may well (e.g., through cancellation effects) ultimately yield a simple final state. Our theoretical framework perfectly accommodates this unidirectional causal relationship.

\subsection{Rigorous Proof of Theorem 4.2 (Process-State Complexity Bound)}
\label{G.2}
\begin{spacing}{1.2} 
    Before proceeding with the formal proof, we must first establish a key physical assumption that underpins Lemma~\ref{G.3}in our argument. This assumption concerns the characteristic nature of the output generated by complex quantum processes and provides the foundation for invoking a central inequality from algorithmic information theory (AIT).

\textbf{Physical Assumption G.1 (Algorithmic Typicality of Final-State Distributions):} For a quantum process governed by a Hamiltonian without special integrable symmetries and evolved for a sufficiently long duration, the final classical probability distribution $P$ obtained upon measurement can be regarded, in the sense of algorithmic information theory, as incompressible. That is, the amount of information required to describe $P$ is of the same order as the total information it contains—quantified by its Shannon entropy $H(P)$.

\textbf{The physical motivation for this assumption stems from quantum chaos and the eigenstate thermalization hypothesis (ETH): }deep quantum evolution rapidly erases any simple structure present in the initial state, resulting in an output distribution that approximates a “pseudorandom” form. On the experimental side, such high-randomness features have been observed in randomized quantum circuit sampling (RCS) and certified quantum randomness experiments—for instance, in the RCS task performed by the Google Sycamore processor\cite{arute2019quantum}. This assumption thus provides solid physical justification for our subsequent analysis of the Kolmogorov complexity of final-state distributions. If future investigations consider integrable or highly symmetric systems, where the distribution $P$ may exhibit compressible structure, then this assumption should not be applied and a separate evaluation of algorithmic information characteristics will be necessary. With this background established, we now proceed to the main proof.
Our proof will proceed in four main steps, progressively building the chain of inequalities that connects $C_{\mathrm{opt}}$ and $\widehat{K}_{\mathrm{KL}}$.

\begin{enumerate}
    \item \textbf{Step 1: From Process Complexity to Process Description Complexity ($C_{\mathrm{opt}}(U) \to K(U)$)}
    We first establish the relationship between the descriptive complexity of a unitary evolution $U$ (i.e., its Kolmogorov complexity $K(U)$) and its optimal computational complexity $C_{\mathrm{opt}}(U)$.

    \textbf{Lemma G.1:} For a quantum computer constructed from a finite universal gate set $G$, the Kolmogorov complexity $K(U)$ of a unitary evolution $U$ is bounded by its optimal computational complexity $C_{\mathrm{opt}}(U)$.\footnote{This proof is rigorous for a finite gate set. For gates with continuous parameters (e.g., arbitrary rotation gates), we assume all gates have been discretized to a precision of $\exp(-\mathrm{poly}(n))$, and the corresponding increase in circuit depth has been absorbed into the definition of $C_{\mathrm{opt}}$, which is consistent with the standard setting of fault-tolerant quantum computation \cite{kitaev2002classical, nielsen2010quantum}.}

    \textbf{Proof:} This is a constructive proof. We can construct a Turing machine program that takes no input and outputs a complete description of $U$ (e.g., its matrix representation to sufficient precision). This program consists of the following parts: \textbf{(i)} a subroutine to generate the value of $C_{\mathrm{opt}}(U)$ itself, with length $K(C_{\mathrm{opt}}(U))$; \textbf{(ii)} a loop that sequentially lists the $C_{\mathrm{opt}}(U)$ gates from the optimal quantum circuit for $U$, where specifying each gate from a set of size $|G|$ requires $\log_2|G|$ bits of information; \textbf{(iii)} a fixed interpreter program of length $O(1)$ that explains the above information and constructs the description of $U$. The total length of this program constitutes an upper bound on $K(U)$. For simplicity, we define a "compilation efficiency" constant $c_G = 1/\log_2|G|$ that depends only on the gate set $G$, and absorb the logarithmic term $K(C_{\mathrm{opt}}(U)) = O(\log C_{\mathrm{opt}})$, yielding the following key inequality:
    \begin{equation}
    K(U) \le (1/c_G) C_{\mathrm{opt}}(U) + O(\log C_{\mathrm{opt}}(U))
    \label{eq:G.1} 
    \end{equation}

    \item \textbf{Step 2: From Process Description Complexity to Result Description Complexity ($K(U) \to K(P)$)}
    Next, we connect the complexity of the process $U$ to the complexity of the classical result it produces (the probability distribution $P$).

    \textbf{Lemma G.2:} If a unitary evolution $U$ acts on a simple initial state $|\psi_0\rangle$ (e.g., $|0...0\rangle$) with Kolmogorov complexity $O(1)$, and a classical probability distribution $P$ is obtained through a measurement scheme described by a program $M$ with Kolmogorov complexity $K(M)$, then the Kolmogorov complexity of $P$, $K(P)$, satisfies the following inequality:
    
    \begin{equation}
    K(P) \le K(U) + K(M) + O(\log K(U))
    \label{eq:G.2}
    \end{equation}

    \textbf{Proof:} This is also a constructive proof based on algorithmic information theory. We can construct a program to output the probability distribution $P$ (to any specified precision). The core parts of this program are: \textbf{(i)} a subroutine to reconstruct the unitary evolution $U$ (length $K(U)$); \textbf{(ii)} a subroutine to reconstruct the measurement scheme $M$ (length $K(M)$); \textbf{(iii)} the main logic, which simulates the quantum evolution $U|\psi_0\rangle$ and calculates the probabilities of all possible measurement outcomes according to Born's rule. Based on a fundamental principle of algorithmic information theory—that the complexity of an output of a computation cannot significantly exceed the sum of the complexities of all its inputs—this inequality holds. We have assumed $K(|\psi_0\rangle) = O(1)$ and absorbed it.

   \item \textbf{Step 3: From Descriptional Complexity to Information Entropy ($K(P) \to H(P)$)}\\ 
   This step connects the descriptional complexity of the result with its statistical complexity. Such a connection does not hold universally, but becomes valid under our foundational physical assumption.

  \textbf{Lemma G.3:} For a probability distribution $P$ that is algorithmically typical in the sense defined by Physical Assumption~G.1, its Kolmogorov complexity $K(P)$ is bounded from below by its Shannon entropy $H(P)$, satisfying the inequality:

\begin{equation}
K(P) \ge H(P) - O(1)
\label{eq:G.3}
\end{equation}

\textit{(Note: The $O(1)$ term depends only on the choice of prefix universal Turing machine and is independent of the system size.)
}

\textbf{Proof:} This inequality is a direct mathematical consequence of Physical Assumption~G.1. The assumption asserts that the distribution $P$ possesses the physical property of being algorithmically typical—i.e., incompressible. Foundational results in algorithmic information theory, particularly the theoretical framework established by the Li-Vitányi Coding Theorem, provide a rigorous connection between the complexity of a typical object and its Shannon entropy.\cite{li2008introduction} Since Assumption~G.1 designates $P$ as an incompressible distribution, and the Li-Vitányi theorem states that such distributions satisfy $K(P) \ge H(P) - O(1)$, the inequality~\eqref{eq:G.3} follows immediately. Therefore, the validity of this lemma rests entirely on the physical plausibility of Assumption~G.1.

\item \textbf{Step 4: Assembling the Proof Chain and Proving the Theorem}
We now assemble the key inequalities established in the first three steps (\eqref{eq:G.1}, \eqref{eq:G.2}, and \eqref{eq:G.3}) to construct a complete inequality chain from the optimal computational complexity $C_{\mathrm{opt}}$ to the Shannon entropy $H(P)$. Our starting point is the lower bound for $C_{\mathrm{opt}}(U)$ given by Lemma G.1. Rearranging inequality \eqref{eq:G.1} gives:
    
    \begin{equation}
    C_{\mathrm{opt}}(U) \ge c_G \left(K(U) - O(\log C_{\mathrm{opt}}(U))\right)
    \end{equation}
    
    To further constrain the terms on the right-hand side, we apply Lemma G.2, which provides a lower bound for $K(U)$ determined by the result complexity $K(P)$. Substituting $K(U) \ge K(P) - K(M) - O(\log K(U))$ from inequality \eqref{eq:G.2} into the above expression, we have:
    
    \begin{equation}
    C_{\mathrm{opt}}(U) \ge c_G \left[ \left(K(P) - K(M) - O(\log K(U))\right) - O(\log C_{\mathrm{opt}}(U)) \right]
    \end{equation}
    
    Next, we apply Lemma G.3, i.e., $K(P) \ge H(P) - O(1)$, to bound the $K(P)$ term. This extends the inequality chain to the Shannon entropy:
    
    \begin{equation}
    C_{\mathrm{opt}}(U) \ge c_G \left[ \left(H(P) - O(1)\right) - K(M) - O(\log K(U)) - O(\log C_{\mathrm{opt}}(U)) \right]
    \end{equation}
    
    By collecting all low-order constant and logarithmic terms, and noting that since $K(U)$ is approximately linear in $C_{\mathrm{opt}}(U)$ (see G.1), the $O(\log K(U))$ term can be absorbed by $O(\log C_{\mathrm{opt}}(U))$, we obtain a concise and powerful theoretical lower bound based on Shannon entropy:
    
    \begin{equation}
    C_{\mathrm{opt}}(U) \ge c_G (H(P) - K(M)) - O(\log C_{\mathrm{opt}}(U))
    \label{eq:G.4}
    \end{equation}
    
    The final step, and the key link between theory and experiment, is to replace the theoretical Shannon entropy $H(P)$ with the experimentally measurable state complexity proxy $\widehat{K}_{\mathrm{KL}}$. We first note the identity $H(P) = K_{\mathrm{KL,true}}$. As described in Sec.~\ref{G.3}, in an experiment with a finite number of samples, there is a bound determined by the statistical error $f(\varepsilon)$ between the measured value $\widehat{K}_{\mathrm{KL}}$ and the true theoretical value $K_{\mathrm{KL,true}}$, i.e., $K_{\mathrm{KL,true}} \ge \widehat{K}_{\mathrm{KL}} - f(\varepsilon)$, which holds with a confidence of at least $1-\delta$. Substituting this relation into inequality \eqref{eq:G.4} and replacing $O(\log C_{\mathrm{opt}}(U))$ with its measurable proxy $O(\log(\widehat{K}_{\mathrm{KL}}))$ for self-consistency, we arrive at the final form of Theorem 4.2—a PAC (Probably Approximately Correct)-style lower bound with statistical guarantees:
    
    \begin{equation}
    C_{\mathrm{opt}} \ge c_G (\widehat{K}_{\mathrm{KL}} - K(M) - f(\varepsilon)) - O(\log(\widehat{K}_{\mathrm{KL}}))
    \label{eq:G.5}
    \end{equation}
    
    This completes the proof of the "Process-State Complexity Bound" theorem.
\end{enumerate}
\end{spacing}

\subsection{Statistical Error Analysis of \texorpdfstring{$\widehat{K}_{\mathrm{KL}}$}{K\_KL}}
\label{G.3}
This section aims to quantify the statistical error introduced when estimating $\widehat{K}_{\mathrm{KL}}$ from a finite number of experimental measurements. The value of $\widehat{K}_{\mathrm{KL}}$ is entirely determined by the estimated probability distribution $\widehat{P}$; therefore, our core task is to analyze how the estimation error in $\widehat{P}$ propagates to $\widehat{K}_{\mathrm{KL}}$. Our analysis is built upon the theoretical guarantees of quantum shadow tomography. According to this theory \cite{huang2020predicting}, by performing $N_{\mathrm{shot}}$ measurements (e.g., using a local Clifford shadow protocol), we can obtain an estimate of the probability distribution, $\widehat{P}$, such that its $\ell_1$-norm distance to the true physical distribution $P$, $||\widehat{P} - P||_1$, is bounded by an error $\varepsilon$ with a confidence of at least $1-\delta$. The required number of measurements, $N_{\mathrm{shot}}$, satisfies:

\begin{equation}
N_{\mathrm{shot}} \ge (c / \varepsilon^2) (\log(M_{\mathrm{out}}) + \log(1/\delta))
\label{eq:G.6}
\end{equation}

where $M_{\mathrm{out}}$ is the number of distinct possible measurement outcomes, and $c$ is a small numerical constant. Next, we use a continuity inequality from information theory to bound the error in $\widehat{K}_{\mathrm{KL}}$. By the definition of $\widehat{K}_{\mathrm{KL}}$, its source of error is its component, the Shannon entropy $H(\widehat{P})$. The well-known Fannes-Audenaert inequality provides a tight bound on the stability of the Shannon entropy \cite{audenaert2007}. An easier-to-use corollary, derived from Pinsker's inequality, is that the Shannon entropy (and the KL divergence) is Lipschitz continuous in its probability distribution argument. When $||\widehat{P} - P||_1 \le \varepsilon$, the absolute error between the estimated value $\widehat{K}_{\mathrm{KL}}$ and its true theoretical value $K_{\mathrm{KL,true}} = H(P)$ satisfies:

\begin{equation}
|\widehat{K}_{\mathrm{KL}} - K_{\mathrm{KL,true}}| \le \varepsilon \log(M_{\mathrm{out}} / \varepsilon) =: f(\varepsilon)
\label{eq:G.7}
\end{equation}

This error bound, $f(\varepsilon)$, behaves well as $\varepsilon \to 0$, ensuring that by increasing the number of measurements $N_{\mathrm{shot}}$ (and thus decreasing $\varepsilon$), we can make the estimated value of $\widehat{K}_{\mathrm{KL}}$ arbitrarily close to its true theoretical value. In the final inequality of Sec.~\ref{G.2} (\eqref{eq:G.5}), this statistical error $f(\varepsilon)$ is explicitly included, yielding a rigorous lower bound with a high-confidence statistical guarantee. In a typical experimental setting, an experimenter can pre-define a target precision (e.g., $\varepsilon = 0.01$) and calculate the required number of measurements according to Eq.~\eqref{eq:G.6} to ensure that the impact of the statistical error $f(\varepsilon)$ on the final result is at a controllable, subordinate level.

\subsection{Details of the Experimental Calibration Protocol}
\label{G.4}
The "compilation efficiency" constant $c_G$ and the "measurement scheme complexity" $K(M)$ introduced in Theorem 4.2, while necessary products of the theoretical derivation, are not free fitting parameters. These constants reflect the intrinsic properties of a specific experimental platform (including its hardware and the chosen universal gate set $G$ and measurement scheme $M$). This section provides a one-time protocol for experimentally calibrating these constants.

\textbf{Protocol Objective:} To calibrate $c_G$ and $K(M)$.

\textbf{Protocol Steps:}
\begin{enumerate}
    \item \textbf{Benchmark Circuit Selection:}
    Select a set of quantum circuits (at least two, the more the better) with varying but known optimal computational complexities, $C_{\mathrm{opt}}$. These $C_{\mathrm{opt}}$ values can come from exact theoretical solutions (e.g., the optimal circuit depth for preparing an $n$-qubit GHZ state is $\log(n)$) or from tight upper bounds provided by powerful classical quantum circuit compilers \cite{amy2013meet}. We denote this set of benchmark circuits and their complexities as $\{(U_i, C_{\mathrm{opt},i})\}$.
    \item \textbf{Execution and Measurement:}
    For each benchmark circuit $U_i$, execute the circuit on your quantum hardware and measure the resulting final state using the shadow tomography protocol described in Sec.~\ref{G.3} to obtain a precise estimate of $\widehat{K}_{\mathrm{KL},i}$ (ensuring the statistical error $f(\varepsilon)$ is sufficiently small).
    \item \textbf{Data Fitting and Parameter Extraction:}
    We now have a set of data points $\{(\widehat{K}_{\mathrm{KL},i}, C_{\mathrm{opt},i})\}$. According to the leading terms of Theorem 4.2 (ignoring lower-order logarithmic terms), these data points should (with high confidence) satisfy the linear relationship:
    
    \begin{equation}
    C_{\mathrm{opt},i} \approx c_G (\widehat{K}_{\mathrm{KL},i} - K(M))
    \end{equation}
    
    \begin{center}
     Rearranging gives:
    \end{center}  
    
    \begin{equation}
    C_{\mathrm{opt},i} \approx c_G \widehat{K}_{\mathrm{KL},i} - c_G K(M)
    \end{equation}
    
    Perform a linear regression fit on these data points, with $C_{\mathrm{opt},i}$ as the y-axis and $\widehat{K}_{\mathrm{KL},i}$ as the x-axis.
    \begin{enumerate} 
        \item The \textbf{slope} of the fit is the experimentally calibrated value of the compilation efficiency constant, $c_G$.
        \item The \textbf{y-intercept} of the fit is equal to $-c_G K(M)$. Dividing the intercept by the already calibrated $-c_G$ yields the calibrated value of the measurement scheme complexity, $K(M)$.
    \end{enumerate}
\end{enumerate}

\textbf{Protocol Notes:}
\begin{enumerate} 
    \item \textbf{One-Time Calibration:} This calibration protocol only needs to be performed once during the lifetime of the experimental platform (or when the gate set or major noise characteristics change significantly). Once $c_G$ and $K(M)$ are calibrated, they can be treated as "fingerprints" of the platform and used for all subsequent performance analyses of unknown complex processes.
    \item \textbf{Properties of $K(M)$:} For a fixed measurement scheme (like local Clifford shadow measurement), $K(M)$ is a slowly growing function of the system size $n$ (i.e., $O(n)$) and primarily acts as an offset in the fit.
    \item \textbf{Significance of $c_G$:} The value of $c_G$ directly quantifies the efficiency with which the hardware compiles information (measured in bits) into physical operations (measured in number of gates), making it a highly valuable, independent hardware benchmark parameter.
\end{enumerate}
Through this protocol, all unknown parameters in Theorem 4.2 are determined, allowing the entire performance-equation framework to be fully closed and applied.

\subsection{On the Relationship Between the Complexity Proxy, Shannon Entropy, and Algorithmic Complexity}
\label{G.5}
This section aims to clarify the formal relationship among the state complexity proxy introduced in this paper ($\widehat{K}_{\mathrm{KL}}$), the Shannon entropy $H(P)$, and the algorithmic (Kolmogorov) complexity $K(P)$, thereby providing a deeper theoretical underpinning for the physical meaning of the proxy.
From a physical perspective, the algorithmic information contained in a probability distribution $P$ is directly related to the richness of its inherent "patterns" or "structure." A uniform distribution $U$ represents complete statistical disorder and contains no structural information that can be exploited by an algorithm for compression. Therefore, any deviation from a uniform distribution implies the existence of a describable pattern in the system. The Kullback-Leibler (KL) divergence, $D_{\mathrm{KL}}(P || U)$, precisely quantifies, in an information-theoretic sense, the "information gain" when updating from a uniform prior $U$ to a posterior distribution $P$ that contains specific information.
This physical picture can be quantified through a rigorous mathematical relationship. We first define the Shannon entropy $H(P) = -\sum_i p_i \log_2(p_i)$ and the KL divergence $D_{\mathrm{KL}}(P || U) = \sum_i p_i \log_2(p_i/u_i)$. For a system with $M_{\mathrm{out}}$ possible outcomes, its uniform distribution is $u_i = 1/M_{\mathrm{out}}$. There exists an exact identity between these two fundamental quantities:

\begin{equation}
H(P) = \log_2(M_{\mathrm{out}}) - D_{\mathrm{KL}}(P || U)
\label{eq:G.9}
\end{equation}

This relationship reveals the physical meaning of $H(P)$: it quantifies the uncertainty of a distribution, and its value is equal to the maximum possible uncertainty ($\log_2(M_{\mathrm{out}})$) minus the "identifiable information" ($D_{\mathrm{KL}}$) contained in the distribution's deviation from uniformity. It is noteworthy that the state complexity proxy defined in this paper, $\widehat{K}_{\mathrm{KL}} = \log_2(M_{\mathrm{out}}) - D_{\mathrm{KL}}(\widehat{P} || U)$, has as its theoretical target precisely the system's Shannon entropy, $H(P)$.
Next, we connect this relationship to a fundamental theorem of algorithmic information theory. This theorem sets a lower bound on the Kolmogorov complexity $K(P)$ of a probability distribution $P$, determined by its Shannon entropy \cite{vitanyi2011information}:

\begin{equation}
K(P) \ge H(P) - O(1)
\label{eq:G.8}
\end{equation}

Substituting Eq.~\eqref{eq:G.9} into Eq.~\eqref{eq:G.8}, we establish a direct relationship between the KL divergence and Kolmogorov complexity:

\begin{equation}
K(P) \ge (\log_2(M_{\mathrm{out}}) - D_{\mathrm{KL}}(P || U)) - O(1)
\label{eq:G.10}
\end{equation}

This inequality proves that the complexity proxy we have proposed ($\widehat{K}_{\mathrm{KL}}$), which is essentially a measurement of the Shannon entropy, does indeed provide an operational and computable lower bound for the incomputable but more fundamental Kolmogorov complexity, $K(P)$.
At a methodological level, obtaining a reliable estimate of the KL divergence from experimental data is key to applying this framework. This typically involves constructing a high-fidelity model of the probability distribution ($\widehat{P}$) from a finite number of samples, using statistical inference or modern machine learning techniques, before proceeding with subsequent calculations. Therefore, this framework not only connects physics with information theory but also provides a concrete entry point for the application of advanced machine learning methods in fundamental physics research.

\bibliographystyle{unsrt}  
\bibliography{references}

\end{document}